\documentclass[journal]{IEEEtran}
\IEEEoverridecommandlockouts

\usepackage{cite}
\usepackage{graphicx}
\usepackage{amsmath}
\usepackage{mdwmath}
\usepackage{amssymb}
\usepackage{mdwtab}
\usepackage{stfloats}
\usepackage[footnotesize]{subfigure}
\usepackage{amsmath,amsthm}
\usepackage{threeparttable}
\usepackage{xcolor}
\usepackage{url}
\usepackage{algorithm}
\usepackage{algpseudocode}
\usepackage{multicol}
\usepackage{multirow}
\usepackage{makecell}
\usepackage{setspace}
\usepackage{bm}
\usepackage{enumitem}
\usepackage{textcase}
\usepackage{balance}
\usepackage{comment}
\usepackage{tikz}
\usepackage{pgfplots}
\pgfplotsset{compat=newest}
\pgfplotsset{plot coordinates/math parser=false}
\newlength\figureheight
\newlength\figurewidth
\allowdisplaybreaks
\algrenewcommand{\algorithmicrequire}{\textbf{Input:}}
\algrenewcommand{\algorithmicensure}{\textbf{Output:}}

\newtheorem{example}{\it  Example}
\newtheorem{theorem}{\it Theorem}
\newtheorem{remark}{\it  Remark}
\newtheorem{lemma}{\it  Lemma}

\newtheorem{definition}{\it  Definition}

\newtheorem{problem}{\it  Problem}
\def\BibTeX{{\rm B\kern-.05em{\sc i\kern-.025em b}\kern-.08em
    T\kern-.1667em\lower.7ex\hbox{E}\kern-.125emX}}

\newcommand*\circled[1]{\tikz[baseline=(char.base)]{
            \node[shape=circle,draw,inner sep=.5pt] (char) {#1};}}

\newcommand{\prob}[1]{\mathrm{Pr}\left\{ #1 \right\}}
\newcommand{\T}{\mathrm{T}}

\usepackage{hyperref}
\hypersetup{
    colorlinks=true,
    linkcolor=red,
    citecolor=red,
    urlcolor=blue,
    breaklinks=true}
\pgfplotsset{compat=1.18}

\usepackage[nolist]{acronym}
\begin{acronym}[ACRONYM]
\acro{AI}{artificial intelligence}
\acro{AoA}{angle-of-arrival}
\acro{AoD}{angle-of-departure}
\acro{BS}{base station}
\acro{BP}{belief propagation}
\acro{CDF}{cumulative density function}
\acro{CFO}{carrier frequency offset}
\acro{CRB}{Cram\'er-Rao bound}
\acro{DA}{data association}
\acro{D-MIMO}{distributed multiple-input multiple-output}
\acro{DL}{downlink}
\acro{EM}{electromagnetic}
\acro{FIM}{Fisher information matrix}
\acro{GDOP}{geometric dilution of precision}
\acro{GNSS}{global navigation satellite systems}
\acro{GPS}{global positioning system}
\acro{IP}{incidence point}
\acro{IQ}{in-phase and quadrature}
\acro{ISAC}{integrated sensing and communication}
\acro{ICI}{inter-carrier interference}
\acro{JCS}{Joint Communication and Sensing}
\acro{JRC}{joint radar and communication}
\acro{JRC2LS}{joint radar communication, computation, localization, and sensing}
\acro{IMU}{inertial measurement unit}
\acro{IOO}{indoor open office}
\acro{IoT}{Internet of Things}
\acro{IRN}{infrastructure reference node}
\acro{KPI}{key performance indicator}
\acro{LoS}{line-of-sight}
\acro{LS}{least-squares}
\acro{MCRB}{misspecified Cram\'er-Rao bound}
\acro{MIMO}{multiple-input multiple-output}
\acro{ML}{maximum likelihood}
\acro{mmWave}{millimeter-wave}
\acro{NLoS}{non-line-of-sight}
\acro{NR}{new radio}
\acro{OFDM}{orthogonal frequency-division multiplexing}
\acro{OTFS}{orthogonal time-frequency-space}
\acro{OEB}{orientation error bound}
\acro{PEB}{position error bound}
\acro{VEB}{velocity error bound}
\acro{PRS}{positioning reference signal}
\acro{QoS}{Quality of Service}
\acro{RAN}{radio access network}
\acro{RAT}{radio access technology}
\acro{RCS}{radar cross section}
\acro{RedCap}{reduced capacity}
\acro{RF}{radio frequency}
\acro{RIS}{reconfigurable intelligent surface}
\acro{RFS}{random finite set}
\acro{RMSE}{root mean squared error}
\acro{RTK}{real-time kinematic}
\acro{RTT}{round-trip-time}
\acro{SLAM}{simultaneous localization and mapping}
\acro{SLAT}{simultaneous localization and tracking}
\acro{SNR}{signal-to-noise ratio}
\acro{ToA}{time-of-arrival}
\acrodefplural{ToA}{times-of-arrival}
\acro{WLS}{weighted least squares}

\acro{TDoA}{time-difference-of-arrival}
\acro{TR}{time-reversal}
\acro{TXRX}[TX/RX]{transmitter/receiver}
\acro{Tx}{transmitter}
\acro{Rx}{receiver}
\acro{UE}{user equipment}
\acro{UL}{uplink}
\acro{UWB}{ultra wideband}
\acro{XL-MIMO}{extra-large MIMO}

\acro{PL}{protection level}
\acro{TIR}{target integrity risk}
\acro{IR}{integrity risk}
\acro{RAIM}{receiver autonomous integrity monitoring}
\acro{1D}{one-dimensional}
\acro{2D}{two-dimensional}
\acro{3D}{three-dimensional}
\acro{SS}{solution separation}
\acro{PMF}{probability mass function}
\acro{PDF}{probability density function}
\acro{GM}{Gaussian mixture}
\acro{CCDF}{complementary cumulative distribution function}
\acro{SVD}{singular value decomposition}
\acro{AD}{automated driving}
\acro{AL}{alert limit}
\acro{LHS}{left-hand side}

\acro{UAV}{unmanned aerial vehicle}
\acro{FDE}{fault detection and exclusion}

\end{acronym}

\begin{document}

\title{Improving 3D Cellular Positioning Integrity with Bayesian RAIM}

\author{Liqin~Ding, 
Gonzalo~Seco-Granados,~\IEEEmembership{Fellow,~IEEE},
Hyowon~Kim,~\IEEEmembership{Member,~IEEE},
Russ~Whiton, 
Erik~G.~Str\"om,~\IEEEmembership{Fellow,~IEEE},
Jonas~Sj\"oberg,~\IEEEmembership{Member,~IEEE},
Henk~Wymeersch~\IEEEmembership{Fellow,~IEEE}
\thanks{This project has been supported in part by the European Union's Horizon 2020 research and innovation programme under grant agreement number 101006664, in part by the Swedish research council (VR) under grant number 2023-03821, in part by the Catalan Government under the ICREA Academia Program and grant 2021SGR00737, in part by the Spanish project PID2023-152820OB-I00 funded by MICIU/AEI/10.13039/501100011033 and ERDF/EU, in part by the National Research Foundation of Korea funded by the Ministry of Science and ICT under grant (RS-2025-25415939), and in part by the 2025 Daejeon RISE Project (DJR2025-12). The authors would like to thank all partners within Hi-Drive for their cooperation and valuable contribution. This paper is extended from \cite{ding2022bayesian}.}
\thanks{L.~Ding was with the Department of Electrical Engineering, Chalmers University of Technology, Gothenburg, Sweden and is now with Ericsson Research, Ericsson AB, Lund, Sweden.}
\thanks{G.~Seco-Granados is with Department of Telecommunications and
Systems Engineering, Universitat Autonoma de Barcelona, Bellaterra, Spain.}
\thanks{H.~Kim is with the Department of Electronics Engineering,
Chungnam National University, Daejeon 34134, South Korea.}
\thanks{R.~Whiton was with Volvo Car Corporation, Gothenburg, Sweden and is now with European Space Agency, Noordwijk, Netherlands.}
\thanks{E.~G.~Str\"om, J.~Sj\"oberg, and H.~Wymeersch are with the Department of Electrical Engineering, Chalmers University of Technology, Gothenburg, Sweden.}
}

\maketitle

\begin{abstract} 
Ensuring positioning integrity amid faulty measurements is crucial for safety-critical applications, making receiver autonomous integrity monitoring (RAIM) indispensable. This paper introduces a Bayesian RAIM algorithm with a streamlined architecture for 3D cellular positioning. Unlike traditional frequentist-type RAIM algorithms, it computes the exact posterior probability density function (PDF) of the position vector as a Gaussian mixture (GM) model using efficient message passing along a factor graph. This Bayesian approach retains all crucial information from the measurements, eliminates the need to discard faulty measurements, and results in tighter protection levels (PLs) in 3D space and 1D/2D subspaces that meet target integrity risk (TIR) requirements. Numerical simulations demonstrate that the Bayesian RAIM algorithm significantly outperforms a baseline algorithm, achieving over $50\%$ PL reduction at a comparable computational cost.
\end{abstract}
\vspace{-4mm}
 
\section{Introduction}

Ensuring integrity in positioning systems is vital for safety-critical applications such as autonomous driving, \acp{UAV}, and industrial automation, where position errors can have catastrophic consequences \cite{zhu2018gnss, isik2020integrity}. Integrity refers to the trustworthiness of the position information and the system’s ability to issue timely warnings when errors exceed acceptable thresholds \cite[Chapter~7.5]{kaplan2006understanding}.  An important measure of integrity is \ac{IR}, which is the probability that the error in the provided position information exceeds an acceptable tolerance without warning the user in a given period of time\cite{zhu2018gnss}. See \cite[Section~II]{zhu2018gnss} for additional integrity-related parameters and definitions.

With advanced cellular technologies like 5G and beyond, cellular network-based positioning has become a promising alternative or complement to \ac{GNSS}, particularly in environments where \ac{GNSS} is unreliable, such as indoors and dense urban areas \cite{de2021convergent,   hammarberg2022architecture}. While cellular positioning has been extensively studied, with numerous algorithms developed to improve positioning accuracy \cite{del2017survey, laoudias2018survey, xue2016ups}, integrity issues remain underexplored. Robust methods are needed to ensure cellular positioning systems meet the stringent integrity requirements of safety-critical applications \cite{bartoletti2021positioning, behravan2022positioning, whiton2022cellular}.

\Ac{RAIM} was originally developed for \ac{GNSS} for aviation navigation and enables receivers to autonomously detect and mitigate faulty measurements \cite{parkinson1988autonomous, brown1996receiver}. Traditional \ac{RAIM} methods typically employ frequentist statistical approaches for \ac{FDE}. These methods rely on measurement redundancy and perform consistency tests either in the measurement domain, such as analyzing measurement residuals \cite{parkinson1988autonomous, brown1996receiver, schroth2008failure, castaldo2014p}, or in the position domain using solution separation testing \cite{blanch2015baseline}. After \ac{FDE}, the \ac{PL} is computed using error bounding methods to ensure that the integrity risk remains within the acceptable \ac{TIR}.

Applying traditional RAIM methods to cellular positioning systems presents significant challenges. On one hand, in typical cellular positioning scenarios, the number of available \acp{BS} is often limited, reducing the measurement redundancy needed for effective \ac{FDE} \cite{rastorgueva2020networking}. Additionally, the challenging propagation environments, such as urban canyons and indoor settings, introduce multipath effects and \ac{NLoS} conditions, complicating the signal measurements \cite{xhafa2021evaluation}. Inexpensive hardware may also introduce clock biases and other signal impairments \cite{koivisto2017joint}. As a result, faults are more prevalent in measurements using cellular signals, and these faults often differ in nature from those in GNSS. Traditional \ac{FDE} methods may struggle to detect and exclude such faults effectively. On the other hand, traditional RAIM methods in GNSS compute separate \acp{PL} for horizontal and vertical directions, which is critical for aviation, while cellular positioning applications require PL computation in full 3D space. Land vehicles operate on varied terrains, and accurate vertical positioning is essential in environments with stacked highways, overpasses, or multi-level structures. UAVs and other airborne vehicles inherently require 3D positioning integrity \cite{xu2020three}. Therefore, extending PL computation to 3D space, as well as to arbitrary 2D or 1D subspaces, is necessary for these applications. Moreover, excluding faulty measurements can discard valuable information, which is particularly detrimental when measurement redundancy is limited. This can lead to conservative \acp{PL}, reducing the availability of the positioning solution, as the system may declare the solution unsafe even when it is acceptable. 

\begin{figure}
\centering
\includegraphics[width=.95\linewidth]{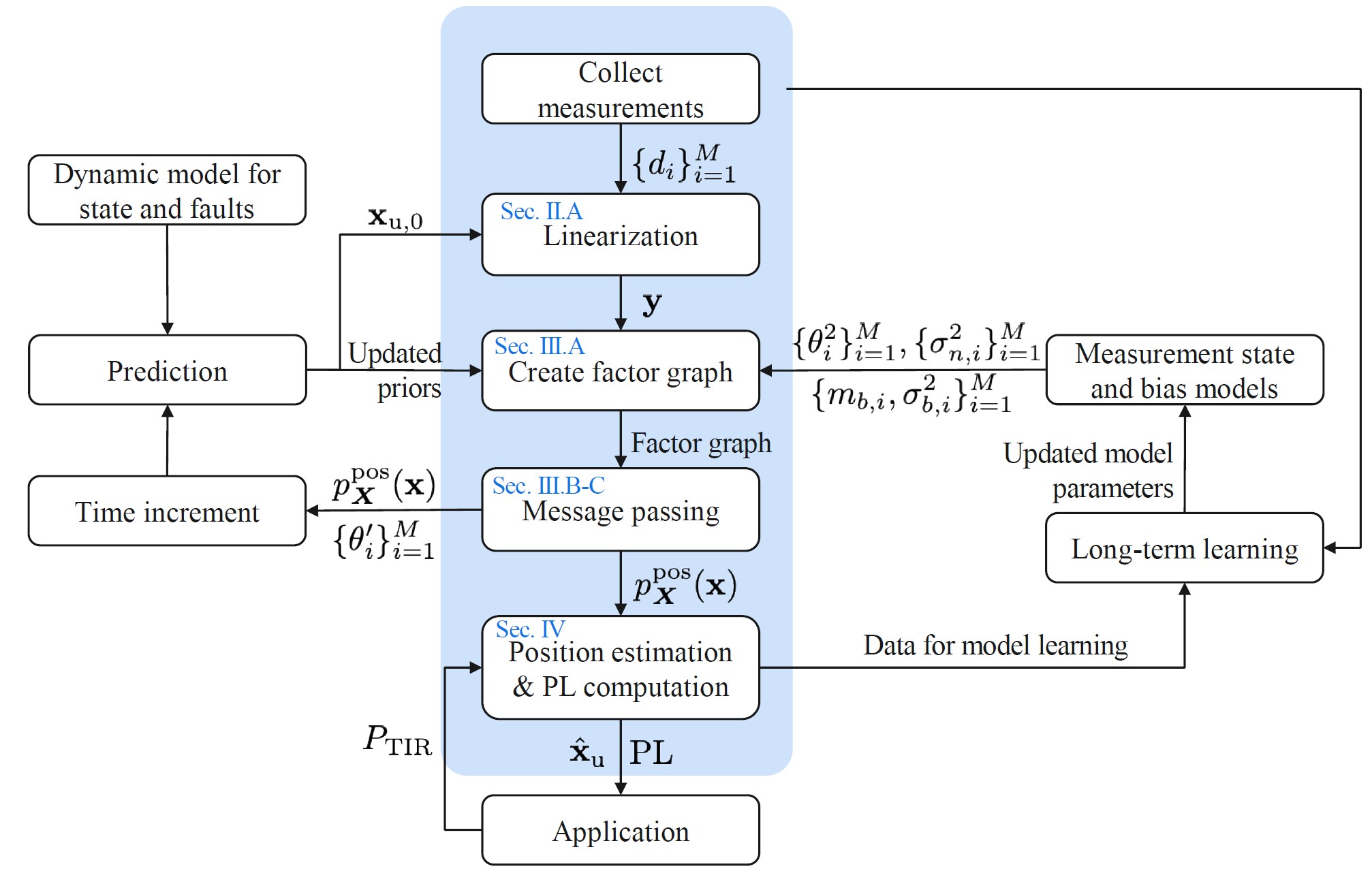}
\caption{Illustration of the Bayesian RAIM framework. A long-term learning module uses hyper-priors to learn fault probabilities and bias models over time. For a single positioning epoch (as highlighted in the blue box and the focus of this paper), these fault probabilities and bias models are considered known. By incorporating dynamic models for both \ac{UE} state and faults, the factor graph approach naturally extends to dynamic scenarios, enabling sequential belief updates for UE and fault tracking. This provides initial UE state estimates and updated fault priors for each positioning epoch.} 
\label{fig:RAIMdiagram}
\vspace{-5mm}
\end{figure}

These challenges can potentially be addressed by the Bayesian RAIM approach \cite{ober2000integrity, pesonen2011framework}, which integrates prior information and all measurements (including potentially faulty ones) into the positioning solution. By modeling faults probabilistically within the estimation process, this approach avoids fault exclusion and yields tighter \acp{PL}, thereby improving availability. However, existing Bayesian RAIM methods often depend on computationally intensive techniques such as particle filters or Monte Carlo sampling \cite{pesonen2011framework, gupta2019particle, gabela2021case}, making them unsuitable for real-time applications.

The Bayesian RAIM framework, after adapting our proposed assumptions and methodologies for \ac{ToA} based cellular positioning, is shown in Fig.~\ref{fig:RAIMdiagram}. It  supports long-term learning of measurement state and bias models through hyper-priors and allows integration of dynamic models for cross-epoch prediction. Focusing on a single epoch (the blue box in Figure~\ref{fig:RAIMdiagram}), in previous work \cite{ding2022bayesian}, we prototyped a Bayesian RAIM algorithm in a 1D setting using efficient message passing on factor graphs. The computed \acp{PL} were significantly tighter than those obtained by a traditional method \cite{blanch2015baseline}. In this paper, we extend this methodology to 3D positioning, addressing new challenges in message passing and PL computation in higher dimensions. Specifically, we focus on (i) accurate computation of \ac{GM} model weights, which is often ignored in previous works \cite{loeliger2004introduction, loeliger2007factor} but is crucial for precise posterior estimation; (ii) efficient handling of degenerate Gaussian densities in message passing, which is essential for computational efficiency and accuracy in multi-dimensional problems; (iii) precise computation of the probability that a Gaussian random vector lies within an arbitrary sphere in $\mathbb{R}^n$ for $n\geq 2$, which allows for application-specific integrity requirements under various scenarios. Our main technical contributions are as follows:

\begin{itemize}
    \item \textbf{Explicit Message Passing Rules for Degenerate Densities:} 
   We introduce efficient computational rules for Gaussian message passing to ensure precise scaling factor calculations for multiple lower-dimensional measurements of a random vector via linear mapping, including rules for the inverse of linear mappings and the product of multiple, potentially degenerate, Gaussian densities. 
    \item \textbf{Tight 2D/3D \ac{PL} Computation:} We develop a numerical integration-based method to accurately assess the probability of a Gaussian-distributed random vector residing within an arbitrary ellipsoid in $\mathbb{R}^n$. This advancement enables precise 2D/3D PL computations based on the \ac{PDF} of the position, whether in Gaussian or \ac{GM} model forms. 
    \item \textbf{Performance Evaluation and Comparison:} Using the above methods, we develop a Bayesian RAIM algorithm for 3D positioning. Monte-Carlo simulations show that our algorithm provides significantly tighter \acp{PL} compared to a baseline RAIM algorithm adapted from \cite{blanch2015baseline}, at a comparable computational cost using a PL overestimation method. With the developed precise PL computation method, it achieves further PL reduction, albeit with increased computational complexity.
\end{itemize}

The remainder of this paper is organized as follows. Section~\ref{sec:Problem} presents the system assumptions and  formally defines the $n$-dimensional \ac{PL}. Section~\ref{sec:BayesianI} describes the construction of the factor graph and details the message passing procedure used to derive the posterior position distribution. Section~\ref{sec:BayesianII} presents different methods for computing the PL. Section~\ref{sec:Baseline} presents a baseline RAIM algorithm that is employed for performance benchmarking. Numerical simulation results are presented in Section~\ref{sec:Numerical:Study}. Finally, Section~\ref{sec:conclusions} offers the conclusions of this work.

\textit{Notations:} 
We use uppercase letters like $X$ for random scalars and boldface uppercase letters like $\boldsymbol{X} = (X_1,\dots,X_n)^{\mathrm T}$ for random vectors. Lowercase letters like $x$ denote deterministic scalars, bold lowercase letters like $\mathbf{x} = (x_1, \dots,x_n)^{\mathrm T}$ denote deterministic vectors, and uppercase sans-serif letters like $\mathsf A$ denote deterministic matrices. We write $\mathsf{A} \succeq 0$ and $\mathsf{A} \succ 0$ for symmetric positive semidefinite and definite matrices, respectively. $\boldsymbol{X} \sim \mathcal{N}(\mathbf{x}; \mathbf{m}, \mathsf{\Sigma})$ indicates that $\boldsymbol{X}$ is a Gaussian vector with mean $\mathbf{m}$ and covariance $\mathsf{\Sigma} \succeq 0$, and $X\sim \mathcal{N}(x; m,\sigma^2)$ indicates that $X$ is a Gaussian variable with mean $m$ and variance $\sigma^2$. 
\vspace{-3mm}

\section{Problem Formulation}
\label{sec:Problem}

\vspace{-1mm}
\subsection{System Assumptions and the Bayesian RAIM Framework}
\label{sec:Problem:a}

We consider a cellular network where $M$ time-synchronized \acp{BS} collaborate in downlink positioning of a \ac{UE}. The positions of the $M$ \acp{BS} in the network coordinate reference system: $\mathbf{x}_i =(x_{i,1},x_{i,2},x_{i,3})^\T \in \mathbb{R}^3$, $i=1,\ldots, M$, are known. In a single positioning epoch, the \acp{BS} send coordinated \acp{PRS} to the \ac{UE} to estimate $M$ \acp{ToA} for the \ac{LoS} paths. These \acp{ToA} convert to pseudoranges as ${d}_i = \Vert \mathbf{x}_i - \mathbf{x}_{\mathrm u} \Vert + x_c + b_i + n_i$, $i=1,\ldots, M$, where $\mathbf{x}_{\mathrm u}=(x_{\mathrm u,1}, x_{\mathrm u,2},x_{\mathrm u,3})^\T \in \mathbb{R}^3$ and $x_c \in \mathbb{R}$ are the \ac{UE}'s 3D position and clock bias, $n_i$ is measurement noise, and $b_i$ represents measurement bias caused by faults.

As is commonly assumed, an initial position estimate $\mathbf{x}_{\mathrm{u},0}$ is available. As illustrated in Fig.~\ref{fig:RAIMdiagram}, it can be predicted based on the dynamic state model. Linearizing $\Vert \mathbf{x}_i - \mathbf{x}_{\mathrm u} \Vert$ around $\mathbf{x}_{\mathrm u,0}$ using a first-order Taylor expansion yields \cite{koivisto2017joint, guvenc2012fundamental, zhu2009simple}
\begin{align}\label{eq:linearization_1}
d_i \approx \Vert \mathbf{x}_i - \mathbf{x}_{\mathrm u,0} \Vert + \mathbf{g}_i^{\mathrm T} (\mathbf{x}_{\mathrm u} - \mathbf{x}_{\mathrm u,0}) + x_c + b_i + n_i,
\end{align}
where $\mathbf{g}_i \triangleq (\mathbf{x}_{\mathrm u,0} - \mathbf{x}_i)/\Vert \mathbf{x}_{\mathrm u,0} - \mathbf{x}_i\Vert$ is the unit vector from $\mathbf{x}_{\mathrm u,0}$ to $\mathbf{x}_i$. Letting $\mathbf{h}_i = [\mathbf{g}_i^\T\; 1]^\T$ and $\mathbf{x} = [\mathbf{x}_\mathrm{u}^\T\; x_c]^\T$, we establish a linear measurement model
\begin{align} \label{eq:linearized-3D-model}
y_i = \mathbf{h}_i^\T \mathbf{x} + b_i + n_i , \quad i=1,\ldots, M,
\end{align}
which serves as an approximation to $d_i-\Vert \mathbf{x}_i - \mathbf{x}_{\mathrm u,0} \Vert  +  \mathbf{g}_i^\T \mathbf{x}_{\mathrm u,0}$.
Arranging the $M$ row vectors ${\mathbf{h}_i^\T}$ into an $M\times 4$ matrix $\mathsf{H}$ and forming ${y_i}$, ${b_i}$, and ${n_i}$ into length-$M$ column vectors $\mathbf{y}$, $\mathbf{b}$, and $\mathbf{n}$, the model can be expressed in matrix form
\begin{align}\label{eq:linearized-3D-model:matrix}
\mathbf{y} = \mathsf{H} \mathbf{x} + \mathbf{b} + \mathbf{n}.
\end{align}
The initial position estimate error $\mathbf{e}_{0} \triangleq \mathbf{x}_{\mathrm u} - \mathbf{x}_{\mathrm u,0}$ introduces approximation errors through the approximation \eqref{eq:linearization_1}, affecting the performance of any method that relies on this model.

We treat $\mathbf{x}\in\mathbb{R}^4$, $b_i$ and $n_i$ as realizations of random vector/variable $\boldsymbol{X} = [X_1, X_2,X_3,X_4]^\T$, $B_i$ and $N_i$, respectively. To indicate the measurement state of $y_i$, we introduce a latent random variable $\Lambda_i$ with realization $\lambda_i$, which follows the Bernoulli \ac{PMF} $p_{\Lambda_i}(\lambda_i)=\theta_i^{\lambda_i}(1-\theta_i)^{(1-\lambda_i)}$ where $0< \theta_i \ll 1$. The value $\Lambda_i=0$ denotes a fault-free measurement with $B_i \sim \mathcal{N}(0, 0)$, and $\Lambda_i=1$ indicates a faulty measurement with $B_i \sim \mathcal{N}(m_{b,i}, \sigma_{b,i}^2)$. Measurement noises are zero-mean Gaussian: $N_i \sim \mathcal{N}(0, \sigma_{n,i}^2)$. 
The random variables $\{\Lambda_i\}$, $\{B_i\}$ and $\{N_i\}$ are considered independent within each set.  By applying hyper-priors under the Bayesian framework, we can continually learn and refine both these models and their parameters over time. In the context of a single positioning epoch, we that assume these models and their parameters are known.

\vspace{-4mm}
\subsection{\texorpdfstring{$n$}{n}-Dimensional Protection Level}
\label{sec:Problem:b}

To measure the integrity of positioning results, \acp{PL} are computed under the specified \ac{TIR} requirements. In the context of 3D positioning, \acp{PL} can be determined for the full 3D space and its 1D or 2D subspaces. We define $n$-dimensional PL ($n=1, 2, 3$) using the 3D  positioning error vector $\mathbf{e}_{\mathrm{3D}} \triangleq \mathbf{x}_{\mathrm u}-\hat{\mathbf{x}}_{\mathrm u} \in \mathbb{R}^3$, where $\hat{\mathbf{x}}_{\mathrm u} \triangleq \hat{\mathbf{x}}_{1:3}$ is the position estimate. Consider two orthogonal unit vectors $\tilde{\mathbf{v}}_1, \tilde{\mathbf{v}}_2 \in \mathbb{R}^3$. The 1D subspace spanned by $\tilde{\mathbf{v}}_1$ is $\mathcal{L}(\tilde{\mathbf{v}}_1) \triangleq \{s \tilde{\mathbf{v}}_1 \mid s \in \mathbb{R}\}$, and the 2D subspace spanned by both vectors is $\mathcal{P}(\tilde{\mathbf{v}}_1, \tilde{\mathbf{v}}_2) \triangleq \{s\tilde{\mathbf{v}}_1 + t\tilde{\mathbf{v}}_2 \mid s, t \in \mathbb{R}\}$. The projections of $\mathbf{e}_{\mathrm{3D}}$ onto these subspaces are  $\mathbf{e}_{\mathrm{1D}} = \tilde{\mathbf{v}}_1^\T\mathbf{e}_{\mathrm{3D}}$ and $\mathbf{e}_{\mathrm{2D}} = (\tilde{\mathbf{v}}_1^\T\mathbf{e}_{\mathrm{3D}},  \tilde{\mathbf{v}}_2^\T\mathbf{e}_{\mathrm{3D}})^\T$.  
\begin{definition}
\label{Def:PL:ForAll}
    The $n$-dimensional \ac{PL} for the position estimate $\hat{\mathbf{x}}_{\mathrm u}$ in $\mathcal{L}(\tilde{\mathbf{v}}_1)$ ($n=1$), $\mathcal{P}(\tilde{\mathbf{v}}_1, \tilde{\mathbf{v}}_2)$ ($n=2$), or $\mathbb{R}^3$ ($n=3$), for a \ac{TIR} of $P_\mathrm{TIR}$, is the minimum distance $r$ such that the actual \ac{IR}, $\prob{ \Vert \mathbf{e}_{n\mathrm{D}} \Vert > r }$, does not exceed $P_\mathrm{TIR}$
    \begin{align}\label{eq:PL:ForAll}
        \mathrm{PL}_{n\mathrm{D}}(P_\mathrm{TIR}) = \min\{ r \mid \prob{ \Vert \mathbf{e}_{n\mathrm{D}} \Vert > r } < P_\mathrm{TIR}\}.
    \end{align}
\end{definition}

Geometrically, $\mathrm{PL}_{n\mathrm{D}}(P_\mathrm{TIR})$ is the radius of the smallest interval, circle, or sphere centered on the estimated position that encompasses $\mathbf{e}_{n\mathrm{D}}$ with a probability of at least $1 - P_\mathrm{TIR}$. For $n=2$ or $3$, $\mathrm{PL}_{n\mathrm{D}}(P_\mathrm{TIR})$ bounds $\mathbf{e}_{n\mathrm{D}}$ in any direction within the specified $n$-dimensional space. If the actual \ac{IR} for a given $r$ can be precisely or approximately computed for 1D subspaces, an overestimate of the 2D/3D PL can be determined using the following lemma.
\begin{lemma}\label{lemma1}
For $n =2$ or $3$, given $\mathbf{e}_{n\mathrm{D}} = (e_1,...,e_n)^\T$ and weights\footnote{Choosing $w_i = 1/n$ for all $i$ is pragmatic when positioning error distributions are similar across coordinate directions. Adjusting these weights can refine the overestimate, leading to tighter results. Conversely, setting $w_i$ close to $0$ can significantly increase  $\mathrm{PL}_{1\mathrm{D},i}$ and consequently enlarge $\mathrm{PL}_{n\mathrm{D}}^U$. 
} 
$\{w_1,...,w_n\}$ such that $0<w_i<1$ for all $i$ and $\sum_{i=1}^n w_i = 1$, if $\{\mathrm{PL}_{1\mathrm{D},1},...,\mathrm{PL}_{1\mathrm{D},n}\}$ are obtained such that 
\begin{align}\label{eq:IR:split}
    \mathrm{PL}_{1\mathrm{D},i} = \min\{ r \mid \prob{ |e_i| > r } < w_i \, P_\mathrm{TIR}\}, 
\end{align}
then $\prob{ \Vert \mathbf{e}_{n\mathrm{D}} \Vert > \mathrm{PL}_{n\mathrm{D}}^U} < P_\mathrm{TIR}$ can be guaranteed, where 
\begin{align}\label{eq:PL:overestimate}
    \mathrm{PL}_{n\mathrm{D}}^U = \Big(\sum\nolimits_{i=1}^n \mathrm{PL}_{1\mathrm{D},i}^2 \Big)^{1/2}.
\end{align}
\begin{proof}
See Appendix~\ref{App:Proof:Lemma1}.
\end{proof}
\end{lemma}

\begin{figure}
\centering
\includegraphics[width=.75\linewidth]{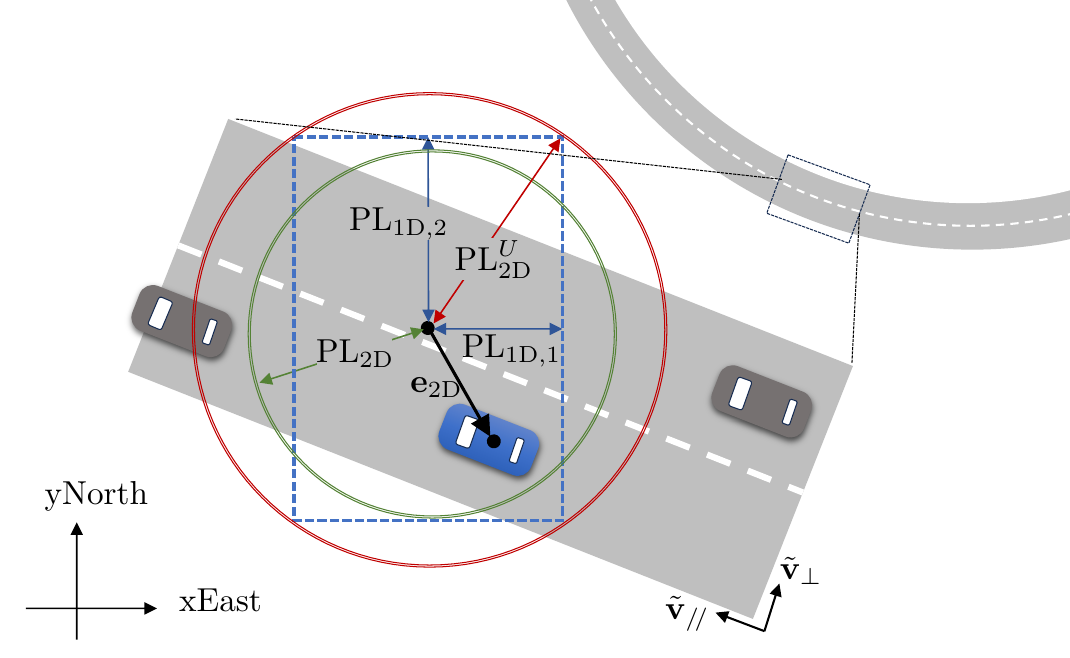} 
	\caption{Illustration of 2D \ac{PL} computation on the horizontal plane.} 
	\vspace{-3mm}
	\label{fig:PL:illustration}
\end{figure}

\begin{example}
Fig.~\ref{fig:PL:illustration} shows a scenario where the \ac{UE} is equipped in a vehicle on a curved road in the horizontal-plane ($\mathrm{x}$-$\mathrm{y}$ plane). To ensure the vehicle remains within its lane, the position error perpendicular to the road ($\tilde{\mathbf{v}}_\perp$ direction) needs to stay within a margin at a confidence level of $1- P_\mathrm{TIR}$. As the vehicle navigates the curve, $\tilde{\mathbf{v}}_\perp$ changes. If $\tilde{\mathbf{v}}_\perp$ is known in real-time, the positioning system can compute $\mathrm{PL}_{1\mathrm{D}}(P_\mathrm{TIR})$ for $\mathcal{L}(\tilde{\mathbf{v}}_\perp)$. Otherwise, it can provide an overestimated 2D PL $\mathrm{PL}_{2\mathrm{D}}^U =(\mathrm{PL}_{1\mathrm{D},1}^2 +\mathrm{PL}_{1\mathrm{D},2}^2)^{1/2}$, where $\mathrm{PL}_{1\mathrm{D},1}$ and $\mathrm{PL}_{1\mathrm{D},2}$ are 1D \acp{PL} computed for $\tilde{\mathbf v}_1 = (1,0,0)^\T$ and $\tilde{\mathbf v}_2 =(0,1,0)^\T$, with $w_1= w_2 = 0.5$. It is conceivable that by slightly decreasing $w_1$ and increasing $w_2$, the blue rectangle can become closer to a square, resulting in a tighter overestimate.
\end{example}

PL requirements vary by application and must align with specific operational demands, see e.g. \cite[Table II]{jing2022integrity}. In Fig.~\ref{fig:PL:illustration}, $\mathrm{PL}_{2D}^U$ might suffice for placing a vehicle on the correct road, suitable for automatic tolling, but it cannot ensure accurate lane positioning, which is crucial for automated driving.

\vspace{-3mm}
\section{Bayesian RAIM Part I: Message Passing}
\label{sec:BayesianI}

The linear Gaussian measurement model \eqref{eq:linearized-3D-model} allows a precise computation of the posterior probability distribution of $\boldsymbol{X}$, $p_{\boldsymbol{X}}^{\mathrm{pos}}(\mathbf{x})$, for each positioning epoch. This section describes the process for obtaining this distribution using message passing on a factor graph. The section concludes with several remarks, including a discussion of computational complexity and an explanation of why measurement exclusion is not recommended.

\vspace{-3mm}
\subsection{Factor Graph Construction and Message Passing Schedule}
\label{sec:BayesianI:0}

\begin{figure}
\centering
\includegraphics[width=.9\linewidth]{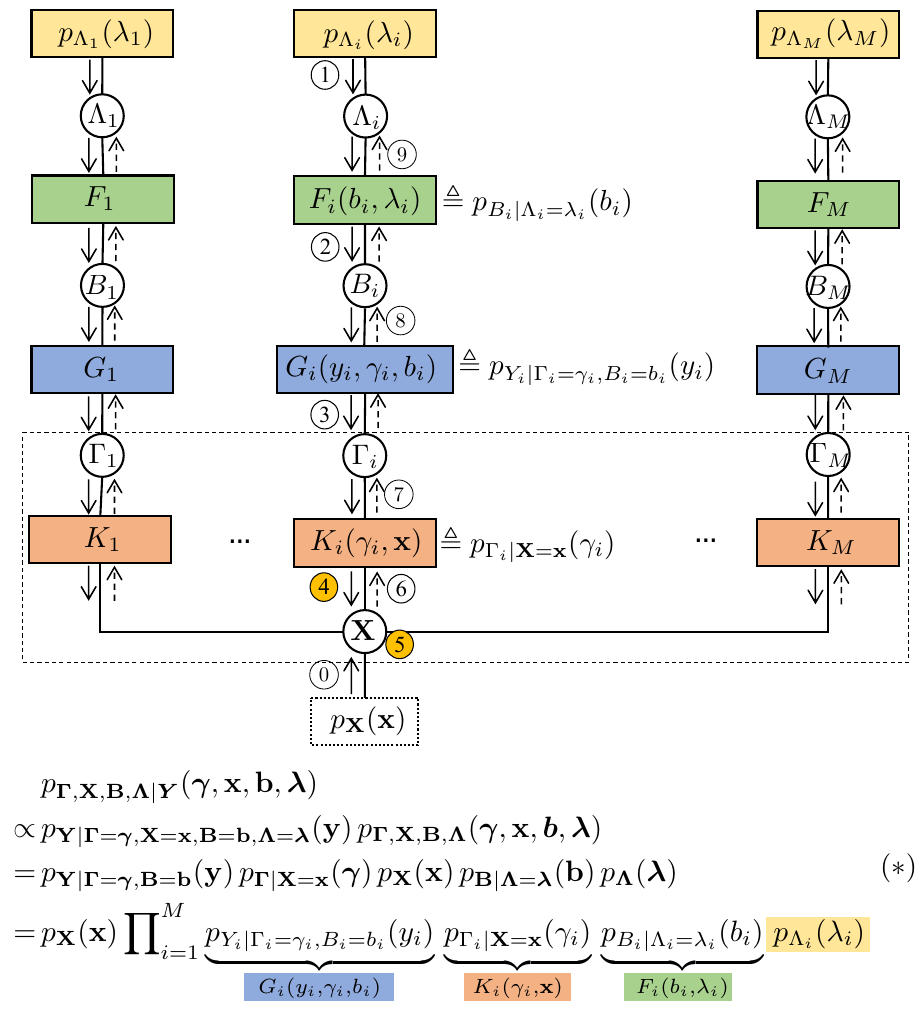} \vspace{-3mm}
\caption{The factor graph used by the Bayesian RAIM algorithm, constructed according to the joint posterior probability factorization given by $(\ast)$. The order of message computation and passing is given by the numbers in circles (shown only on the $i$th branch, but is the same for all branches), while the arrows indicate the passing direction.} 
\label{fig:FactorGraph:longer}
\vspace{-5mm}
\end{figure}

First, we introduce an auxiliary random variable $\Gamma_i = \mathbf{h}_i^{\mathrm T} \boldsymbol{X}$ for $i=1,\ldots, M$, so that we can rewrite the measurement model \eqref{eq:linearized-3D-model} as $y_i = \gamma_i + b_i + n_i$, where $\gamma_i$ is a realization of $\Gamma_i$. We form the sets of random variables $\{Y_i\}$, $\{B_i\}$, $\{\Gamma_i\}$, and $\{\Lambda_i\}$ into random vectors $\boldsymbol{Y}$, $\boldsymbol{B}$, $\boldsymbol{\Gamma}$, and $\boldsymbol{\Lambda}$, and their realizations into vectors $\mathbf{y}$, $\mathbf{b}$, $\boldsymbol{\gamma}$, and $\boldsymbol{\lambda}$. As shown in Fig.~\ref{fig:FactorGraph:longer}, we factorize the joint posterior probability of $\boldsymbol{\Gamma}$, $\boldsymbol{X}$, $\boldsymbol{B}$, and $\boldsymbol{\Lambda}$ in equation $(\ast)$ following the assumptions in Section \ref{sec:Problem:a}, and form a cycle-free factor graph with $M$ branches to represent the factorization. Each term in equation $(\ast)$ corresponds to a factor (function) node in the graph. For ease of description, we define
\begin{subequations}
\begin{align}\label{eq:Gi}
    G_i(y_i,\gamma_i,b_i) &\triangleq p_{Y_i\mid \Gamma_i=\gamma_i, B_i=b_i}(y_i), 
\end{align} 
\begin{align}\label{eq:Ki}
    K_i(\gamma_i,\mathbf{x}) \triangleq p_{\Gamma_i \mid \boldsymbol{X} = \mathbf{x}}(\gamma_i) = \delta(\gamma_i - \mathbf{h}_i^{\mathrm T} \mathbf{x}),
\end{align}
\begin{align}\label{eq:Fi}
  F_i(b_i,\lambda_i) &\triangleq p_{B_i \mid \Lambda_i =\lambda_i}(b_i). 
\end{align} 
\end{subequations} 
In equation~$(\ast)$, $p_{\mathbf X}(\mathbf x)$ represents the prior \ac{PDF} of $\mathbf X$. Since we are focusing on a single epoch, we will omit $p_{\mathbf X}(\mathbf x)$ from message passing. 

Applying the sum-product rule \cite{kschischang2001factor} on this cycle-free factor graph leads to a straightforward message passing schedule: Messages are passed from variable node $\Lambda_i$ to variable node $\mathbf{X}$ in steps \circled{1}-\circled{4} along all $M$ branches in parallel. Then in step \circled{5}, variable node $\mathbf{X}$ computes the product of all received messages, which after normalization is the posterior \ac{PDF} of $\mathbf{X}$. These steps are sufficient for position estimation and PL computation of the current positioning epoch. Messages can be further passed back to variable node $\Lambda_i$ in steps \circled{6}-\circled{9} along each branch to obtain the posterior \ac{PMF} of $\Lambda_i$.\footnote{In a real-life positioning system, measurement states in consecutive epochs are likely correlated, making it reasonable to update $p_{\Lambda_i}(\lambda_i)$ for the next positioning epoch using the obtained posterior \ac{PMF}. However, as this is beyond the scope of this paper, so steps \circled{6}-\circled{9} are not discussed further.}

\vspace{-3mm}
\subsection{General Gaussian Message Computation Rules}
\label{sec:BayesianI:Rules}

Message computation in steps \circled{4}-\circled{6} is challenging because the messages consist of \ac{GM} models, i.e. weighted sums of Gaussians. They require precise computation of weights, and the Gaussian densities may be degenerate with rank-deficient covariance matrices. 
Specifically, step \circled{4} involves inferring a message of $\boldsymbol{X}$ from a lower-dimensional subspace GM model of $Y_i$, which yields degenerate Gaussian densities. Steps \circled{5} and \circled{6} involve computing the product of multiple such messages. Appendix~\ref{App:GMP}, rigorously addresses the computational rules for these steps in the following general problem setting. 

\begin{problem}
\label{prob:1}
    For $i=1,\ldots,K$, given a linear mapping $\boldsymbol{Y}_i = \mathsf A_i \boldsymbol{X}$, where $\boldsymbol{X}\in\mathbb{R}^n$ and $\boldsymbol{Y}_i\in\mathbb{R}^{m_i}$ are Gaussian random vectors, each matrix $\mathsf{A}_i \in \mathbb{R}^{m_i\times n}$ is full-rank\footnote{This full-rank assumption does not lose generality. If $\mathsf A$ is rank-deficient, i.e., $k = \mathrm{rank}(\mathsf A) < \min(n,m)$, a (rank) decomposition $\mathsf A = \mathsf{BC}$ can be obtained (e.g., using compact \ac{SVD}) where $\mathsf{B} \in \mathbb{R}^{m\times k}$ and $\mathsf{C} \in \mathbb{R}^{k\times n}$ have full rank ($=k$). Thus, $\boldsymbol{Y} = \mathsf{A}\boldsymbol{X} =\mathsf{B}(\mathsf{C}\boldsymbol{X})$ can be treated as two consecutive linear mapping via full-rank matrices.}, with $\mathrm{rank}(\mathsf A_i) = m_i \leq n$. The objective is to infer the \ac{PDF} of $\boldsymbol{X}$ from the known non-degenerate \ac{PDF} of $\boldsymbol{Y}_i$ for all $i$, and then compute the product of all inferred \acp{PDF} of $\boldsymbol{X}$. 
\end{problem}

By letting $\mathsf{A}_i = \mathbf{h}_i^{\mathrm T}$, our problem becomes a special case of this general problem.  A general Gaussian distribution for a random vector $\boldsymbol{X} \in\mathbb{R}^n$ is characterized by a symmetric positive semi-definite covariance matrix $\mathsf{\Sigma} \succeq 0$ and a mean vector $\mathbf{m} \in \mathbb R^n$, or equivalently, by $\mathsf{V} \triangleq \mathsf{\Sigma}^{+}$ (the pseudo-inverse of $\mathsf{\Sigma}$) and $\mathbf{u} \triangleq \mathsf{V} \mathbf{m}$. Notably, $\mathrm{rank}(\mathsf{\Sigma}) = \mathrm{rank}(\mathsf{V})\leq n$, with the $<$ relation occurring when the distribution is degenerate. The \ac{PDF} expressed using $(\mathbf{m}, \mathsf{\Sigma})$ is denoted $f_{\boldsymbol{X}}({\mathbf{x}}; \mathbf{m}, \mathsf{\Sigma})$ (see \eqref{eq:pdf_multivariate_normal}), and when using $(\mathbf{u}, \mathsf{V})$, it is denoted $f^{\mathrm{E}}_{\boldsymbol{X}}({\mathbf{x}}; \mathbf{u}, \mathsf{V})$ (see \eqref{eq:pdf_multivariate_normal3}). Moreover, 
\begin{align}\label{eq:alpha}
\alpha_{\boldsymbol{X}} \triangleq - \frac{1}{2}  \mathbf{m}^\T \mathsf{\Sigma}^{+} \mathbf{m} \equiv - \frac{1}{2}  \mathbf{u}^\T \mathsf{V}^{+} \mathbf{u}
\end{align} 
is defined to simplify expressions. The computation rules are summarized in Lemma~\ref{lemma:InverseA}, Lemma~\ref{lemma:Gaussian:product}, and Remark~\ref{remark:final:rule}.

\begin{lemma}[Inverse of linear mapping]
\label{lemma:InverseA}
Consider a linear mapping $\boldsymbol{Y} = \mathsf{A}\boldsymbol{X}$, where $\boldsymbol{X}\in\mathbb{R}^n$ and $\boldsymbol{Y}\in\mathbb{R}^m$ are Gaussian random vectors, $\mathsf{A} \in \mathbb{R}^{m\times n}$, and $\mathrm{rank}(\mathsf A) = m\leq n$. Given a message of $\boldsymbol{Y}$: ${\mu}_{\boldsymbol{Y}}(\mathbf{y})=f_{\boldsymbol{Y}}^{\mathrm{E}}({\mathbf{y}}; \mathbf{u}_{\boldsymbol{Y}}, \mathsf{V}_{\boldsymbol{Y}})$ with $\mathsf{V}_{\boldsymbol{Y}} \succ 0$, the inferred message of $\boldsymbol{X}$ following the sum-product rule is given by  
\begin{subequations}
\begin{align}
    {\mu}_{\boldsymbol{X}}(\mathbf{x}) 
    &= \int \delta(\mathbf{y}- \mathsf{A} \mathbf{x}) {\mu}_{\boldsymbol{Y}} (\mathbf{y}) \mathrm{d}\mathbf{y} = f_{\boldsymbol{Y}}^{\mathrm{E}}({\mathsf{A}\mathbf{x}}; \mathbf{u}_{\boldsymbol{Y}}, \mathsf{V}_{\boldsymbol{Y}}), \nonumber \\
    &= s^{\boldsymbol{X}}_{\boldsymbol{Y}} \cdot f_{\boldsymbol{X}}^{\mathrm{E}}({\mathbf{x}}; \mathbf{u}_{\boldsymbol{X}}, \mathsf{V}_{\boldsymbol{X}}), \label{eq:MM:backward1}
\end{align}
where 
\begin{align}\label{eq:MM:backward2}
\mathbf{u}_{\boldsymbol{X}} = {\mathsf{A}}^\T\mathbf{u}_{\boldsymbol{Y}}, \quad 
\mathsf{V}_{\boldsymbol{X}} = {\mathsf{A}}^\T\mathsf{V}_{\boldsymbol{Y}} \mathsf{A}, 
\end{align}
and the scaling factor $s^{\boldsymbol{X}}_{\boldsymbol{Y}}$ is given by
\begin{align}\label{eq:MM:backward3}
    s^{\boldsymbol{X}}_{\boldsymbol{Y}} = 
  (  |{\mathsf{V}}_{\boldsymbol{Y}}|/|\mathsf{V}_{\boldsymbol{X}}|_+)^{1/2} . 
\end{align}
\end{subequations}
\end{lemma}
\begin{proof}
    See Appendix~\ref{sec:GMP:a}. 
\end{proof}
For $m=n$, $\mathsf{V}_{\boldsymbol{Y}}\succ 0$ ensures $\mathsf{V}_{\boldsymbol{X}} \succ 0$ (since $\mathsf{A}$ is full-rank), making the inverse mapping of $\boldsymbol{Y} = \mathsf{A}\boldsymbol{X}$ unique and straightforward to compute. In this case, \eqref{eq:MM:backward1}-\eqref{eq:MM:backward3} can be directly obtained following \cite[Eq.~(356), (357)]{petersen2008matrix}.

\begin{lemma}[Product of multiple Gaussian densities]
\label{lemma:Gaussian:product}
    The product of $K$ Gaussian \acp{PDF} for $\boldsymbol{X}$: $f^{\mathrm{E}}_{\boldsymbol{X}}({\mathbf{x}}; \mathbf{u}_{\boldsymbol{X}_i}, \mathsf{V}_{\boldsymbol{X}_i})$, where $\mathrm{rank}( \mathsf{V}_{\boldsymbol{X}_i}) =k_{i}$, for $i=1,\ldots, K$, is given by 
\begin{subequations}
    \begin{align}\label{eq:PMG:1}
        \prod_{i=1}^K f^{\mathrm{E}}_{\boldsymbol{X}}({\mathbf{x}}; \mathbf{u}_{\boldsymbol{X}_i}, \mathsf{V}_{\boldsymbol{X}_i})  = s_{\boldsymbol{X}_{1:K}} \cdot f^{\mathrm{E}}_{\boldsymbol{X}}({\mathbf{x}}; \mathbf{u}_{\boldsymbol{X}}, \mathsf{V}_{\boldsymbol{X}}) 
    \end{align}
where 
\begin{align}\label{eq:PMG:2}
    \mathbf{u}_{\boldsymbol{X}} = \sum\nolimits_{i=1}^K \mathbf{u}_{\boldsymbol{X}_i} , \quad  \mathsf{V}_{\boldsymbol{X}} = \sum\nolimits_{i=1}^K \mathsf{V}_{\boldsymbol{X}_i}, 
\end{align}
and the scaling factor $s_{\boldsymbol{X}_{1:K}}$ is, with $k=\mathrm{rank}( \mathsf{V}_{\boldsymbol{X}})$, 
\begin{align}
\label{eq:PMG:3}
    s_{\boldsymbol{X}_{1:K}} = \frac{\prod_{i=1}^K |\mathsf{V}_{\boldsymbol{X}_i}|_+^{1/2}}{(2\pi)^{(\sum_{i=1}^K k_{i} -k)/2}{|\mathsf{V}_{\boldsymbol{X}}|_+^{1/2}}}   
  \exp\Big(\!-\alpha_{\boldsymbol{X}}+\sum_{i=1}^K\alpha_{\boldsymbol{X}_i} \Big).
\end{align}
\end{subequations}
\end{lemma}

\begin{proof}
    See Appendix~\ref{sec:GMP:b}. 
\end{proof}

\begin{remark}
\label{remark:final:rule}
The objective of Problem~\ref{prob:1} can be achieved by applying Lemma~\ref{lemma:InverseA} and Lemma~\ref{lemma:Gaussian:product}, with the scaling factor computations \eqref{eq:MM:backward3} and \eqref{eq:PMG:3} replaced respectively by  
\begin{align}\label{eq:MM:backward3:replacement}
    s^{\boldsymbol{X}}_{\boldsymbol{Y}} =  |{\mathsf{V}}_{\boldsymbol{Y}}|_+^{1/2} 
\end{align}
and 
\begin{align}\label{eq:PMG:3:replacement}
    s_{\boldsymbol{X}_{1:K}} =  \frac{\exp (- \alpha_{\boldsymbol{X}}+\sum_{i=1}^K\alpha_{\boldsymbol{X}_i} )}{(2\pi)^{(\sum_{i=1}^K k_{i} -k)/2}{|\mathsf{V}_{\boldsymbol{X}}|_+^{1/2}}}.
\end{align}
These replacements avoid the computation of $|\mathsf{V}_{\boldsymbol{X}_i}|_+$ while preserving the results. 
\end{remark}

\vspace{-3mm}
\subsection{Message Passing Algorithm}
\label{sec:BayesianI:a}

Applying the above computation rules, a detailed description of steps \circled{1}-\circled{5}, and a brief description of the optional steps \circled{6}-\circled{9}, are given in Appendix~\ref{App:Message:Passing}. Steps \circled{8} and \circled{9} are identical to steps \circled{5} and \circled{6} in \cite[Section~IV.A]{ding2022bayesian} where more details can be found. Additionally, a few remarks should be made regarding step \circled{5}. 
\begin{itemize}
    \item First, when $M\geq 4$ and the measurement vectors $\mathbf{h}_1,\mathbf{h}_2,\ldots, \mathbf{h}_M$ are linearly independent, the matrices ${\mathsf{V}}\!\,_{\boldsymbol{X}}^{(l)}$, $l=1,\ldots,L$, calculated in \eqref{eq:MP:step5:a}, are guaranteed to be positive definite. Consequently, the pseudo-inverse and pseudo-determinant of ${\mathsf{V}}\!\,_{\boldsymbol{X}}^{(l)}$ simplify to the regular inverse and determinant, making the computation of $|\mathsf{V}^{(l)}_{\boldsymbol{X}}|_+ \equiv |\mathsf{V}^{(l)}_{\boldsymbol{X}}|$ in \eqref{eq:MP:step5:c} straightforward. Moreover, the covariance matrix $\mathsf{\Sigma}_{\boldsymbol{X}}^{(l)} = (\mathsf{V}^{(l)}_{\boldsymbol{X}})^+ \equiv (\mathsf{V}^{(l)}_{\boldsymbol{X}})^{-1}$ and mean vector $\mathbf{m}^{(l)}_{\boldsymbol{X}} = (\mathsf{V}^{(l)}_{\boldsymbol{X}})^{-1}\mathbf{u}$ can also be easily computed for all $l$. Thus, we can rewrite \eqref{eq:MP:step5} using Gaussian \acp{PDF} $f_{\boldsymbol{X}}(\mathbf{x};  {\mathbf{m}}^{(l)}_{\boldsymbol{X} } , {\mathsf{\Sigma}}\!\,^{(l)}_{\boldsymbol{X}})$ instead. 
    \item Second, if a prior \ac{PDF} $p_{\boldsymbol{X}}(\mathbf{x})$ is available, it should be multiplied to $\mu_{\boldsymbol{X}} (\mathbf{x})$. This computation can be easily performed if $p_{\boldsymbol{X}}(\mathbf{x})$ is a Gaussian distribution or \ac{GM}. 
    \item Third, the weights $\{w^{(l)}_{\boldsymbol{X}}\}$ given by \eqref{eq:MP:step5:b} do not necessarily sum to one, so a normalization step is required. We reuse $w^{(l)}_{\boldsymbol{X}}$ for the normalized weight to avoid introducing extra symbols. The exact posterior \ac{PDF} of $\boldsymbol{X}$ can be obtained from $\mu_{\boldsymbol{X}} (\mathbf{x})$ as follows: 
\begin{align}\label{eq:pdf_X_pos}
    p_{\boldsymbol{X}}^{\mathrm{pos}}(\mathbf{x}) = \sum\nolimits_{l=1}^{L}  w^{(l)}_{\boldsymbol{X}} \, f_{\boldsymbol{X}}(\mathbf{x};  {\mathbf{m}}^{(l)}_{\boldsymbol{X} } , {\mathsf{\Sigma}}\!\,^{(l)}_{\boldsymbol{X}} ).
\end{align} 
\end{itemize}

\subsubsection{Complexity discussion}
\label{remark:complexity:bayesianI}
The computational complexity of the message passing process scales with $M$. The main computational load comes from step \circled{4}, which is executed in parallel for each measurement, and step \circled{5}, which is performed once. Among the optional steps, step \circled{6} is the most demanding, followed by step \circled{7}. However, most operations in these steps are simple vector or matrix additions and multiplications, and many intermediate results can be reused between steps \circled{5} and \circled{6}.

\subsubsection{Measurement exclusion not recommended}

In \cite{ding2022bayesian}, we proposed excluding a measurement $y_i$ if its posterior fault probability exceeds a threshold $\theta_\mathrm{T}$ and using the posterior \ac{PDF} of $\boldsymbol{X}$ computed with the remaining measurements, i.e., 
$p_{\boldsymbol{X}}^{\mathrm{ex}}(\mathbf{x}) \propto \prod_{j\in \mathcal{I}_{\mathrm{ex}}^c}  \mu_{K_j \rightarrow \boldsymbol{X}} (\mathbf x)$, where $\mathcal{I}_{\mathrm{ex}}^c \triangleq\{1\leq i\leq M: \theta_i'\leq\theta_\mathrm{T}\}$, instead of \eqref{eq:pdf_X_pos} for \ac{PL} computation. However, this approach can compromise integrity requirements. Indeed, measurement exclusion is unnecessary since all information about the UE position and measurement states is captured in $p_{\boldsymbol{X}}^{\mathrm{pos}}(\mathbf{x}) \propto \prod_{j=1}^M  \mu_{K_j \rightarrow \boldsymbol{X}} (\mathbf x)$. In fact, if exclusion is performed, the conditional pointwise mutual information (C-PMI) between $\boldsymbol{X}$ and the excluded measurements, given the remaining measurements, must be considered to maintain integrity requirements.  
To elaborate, we revisit the nature of the problem: In each positioning epoch, the position $\mathbf{x}\in \mathbb{R}^n$ and the measurement $\mathbf{y}\in \mathbb{R}^M$ are generated according to their joint probability \ac{PDF} $p_{\boldsymbol{X,Y}}(\mathbf{x},\mathbf{y})$. Based on the posterior \ac{PDF} $p_{\boldsymbol{X}\mid \boldsymbol{Y} =\mathbf{y}}(\mathbf{x})$, a $n$-dimensional ball $\mathcal B(\mathbf{y})$, centered at $\hat{\mathbf{x}}$ with radius $\mathrm{PL}$, is determined to satisfy the requirement $\int p_{\boldsymbol{X}\mid \boldsymbol{Y} =\mathbf{y}}(\mathbf{x}) I(\mathbf{x}\in \mathcal B(\mathbf{y})) \mathrm{d} \mathbf{x} \geq 1 - P_{\mathrm{TIR}}$, where $I(\cdot)$ stands for the indicator function. The integrity probability, which is the reciprocal of the integrity risk and can be expressed as 
\begin{align*}
    P_{\mathrm{I}} &= \int \int p_{\boldsymbol{X,Y}}(\mathbf{x},\mathbf{y}) I(\mathbf{x}\in \mathcal B(\mathbf{y})) \mathrm{d} \mathbf{x} \mathrm{d} \mathbf{y} \\
    & = \int p_{\boldsymbol{Y}}(\mathbf{y}) \Big( \int p_{\boldsymbol{X}\mid \boldsymbol{Y} =\mathbf{y}}(\mathbf{x}) I(\mathbf{x}\in \mathcal B(\mathbf{y})) \mathrm{d} \mathbf{x}\Big) \mathrm{d} \mathbf{y}  
\end{align*} 
is ensured to be $\geq 1 - P_{\mathrm{TIR}}$. 

As an example, let us consider that for a set of measurements, we have a rule to exclude the first measurement $y_1$. Then, in the realizations where the rule establishes that measurement $y_1$ has to be removed,  a $n$-dimensional ball $\bar{\mathcal B}(\bar{\mathbf{y}})$ is obtained based on the the posterior \ac{PDF} $p_{\boldsymbol{X}\mid \overline{\boldsymbol{Y}} =\bar{\mathbf{y}} }(\mathbf{x})$, where $\bar{\mathbf{y}}\triangleq [y_2,\ldots, y_M]^\T$. This exclusion rule corresponds to region $\mathcal{D}\in \mathbb{R}^M$: If $\mathbf{y}\in \mathcal{D}$, $y_1$ is excluded. Letting $\mathcal{D}^\mathrm{c} \triangleq \mathbb{R}^M\setminus \mathcal{D}$, the integrity probability decomposes as $P_{\mathrm{I}} = P_{\mathrm{I},1} + P_{\mathrm{I},2}$ where $P_{\mathrm{I},1} = \int_{\mathbf{y}\in \mathcal{D}^\mathrm{c}} p_{\boldsymbol{Y}}(\mathbf{y}) \left( \int p_{\boldsymbol{X}\mid \boldsymbol{Y} =\mathbf{y}}(\mathbf{x}) I(\mathbf{x}\in \mathcal B(\mathbf{y})) \mathrm{d} \mathbf{x}\right) \mathrm{d} \mathbf{y}$ and 
$P_{\mathrm{I},2} = \int_{\mathbf{y}\in \mathcal{D}} p_{\boldsymbol{Y}}(\mathbf{y}) \left( \int p_{\boldsymbol{X}\mid \boldsymbol{Y} =\mathbf{y}}(\mathbf{x}) I(\mathbf{x}\in \bar{\mathcal B}(\bar{\mathbf{y}})) \mathrm{d} \mathbf{x}\right) \mathrm{d} \mathbf{y}$. Since 
\begin{align*}
     &\int p_{\boldsymbol{X}\mid \boldsymbol{Y} =\mathbf{y}}(\mathbf{x}) I(\mathbf{x}\in \bar{\mathcal B}(\bar{\mathbf{y}})) \mathrm{d} \mathbf{x}  \nonumber \\
    = &\int \underbrace{\frac{p_{\boldsymbol{X}, Y_1 \mid \overline{\boldsymbol{Y}} =\bar{\mathbf{y}}}(\mathbf{x},y_1)}{p_{\boldsymbol{X} \mid \overline{\boldsymbol{Y}} =\bar{\mathbf{y}}}(\mathbf{x})
    p_{Y_1 \mid \overline{\boldsymbol{Y}} =\bar{\mathbf{y}}}(y_1)} }_{\mathrm{C-PMI}}
    p_{\boldsymbol{X}\mid \overline{\boldsymbol{Y}} =\bar{\mathbf{y}}}(\mathbf{x}) I(x\in  \bar{\mathcal{B}}(\bar{\mathbf{y}}) ) \mathrm{d} \mathbf{x},
\end{align*}
and the C-PMI term, which measures the dependence between $\mathbf{X}$ and $Y_1$ given $\overline{\boldsymbol{Y}} =\bar{\mathbf{y}}$, can be either greater or less than $1$, $P_{\mathrm{I},1} + P_{\mathrm{I},2} \geq 1 - P_{\mathrm{TIR}}$ is \emph{generally not guaranteed}\footnote{Actually, it was observed also experimentally that the inequality is violated when a exclusion rule is implemented.} 
given that $\bar{\mathcal B}(\bar{\mathbf{y}})$ meets the requirement $\int p_{\boldsymbol{X}\mid \overline{\boldsymbol{Y}} =\bar{\mathbf{y}} }(\mathbf{x}) I(\mathbf{x}\in \bar{\mathcal B}(\bar{\mathbf{y}})) \mathrm{d} \mathbf{x} \geq 1 - P_{\mathrm{TIR}}$. 
Intuitively, the issue is that we use the information in $y_1, \ldots, y_M$ to exclude $y_1$ but then use $p_{\boldsymbol{X}\mid \overline{\boldsymbol{Y}} =\bar{\mathbf{y}} }(\mathbf{x})$ for PL computation as if $y_1$ never existed, but $p_{\boldsymbol{X}\mid \overline{\boldsymbol{Y}} =\bar{\mathbf{y}} }(\mathbf{x})$ is not the actual posterior PDF of $\mathbf{X}$ because $y_1$ existed and the whole vector $\mathbf{y}$ was such that it triggered the exclusion rule.

\vspace{-3mm}
\section{Bayesian RAIM  Part II: PL Computation}
\label{sec:BayesianII}

This section introduces the exact and overestimate methods for the $n$-dimensional \ac{PL} based on the posterior \ac{PDF} \eqref{eq:pdf_X_pos}. 

\vspace{-2mm}
\subsection{Exact 1D PL and 2D/3D PL Overestimates}
\label{sec:BayesianII:a}

Since this work does not focus on position estimation methods, the Bayesian RAIM algorithm simply computes the weighted mean, $\overline{\mathbf{m}}_{\boldsymbol{X}} \triangleq \sum\nolimits_{l=1}^L  w^{(l)}_{\boldsymbol{X}}  \mathbf{m}^{(l)}_{\boldsymbol{X}}$, as the estimate of $\mathbf{x}$. Thus, the 3D position estimate is given by $ \hat{\mathbf{x}}_\mathrm{u} 
    = [\overline{\mathbf{m}}_{\boldsymbol{X}} ]_{1:3} 
    = \sum\nolimits_{l=1}^L  w^{(l)}_{\boldsymbol{X}}  [\mathbf{m}^{(l)}_{\boldsymbol{X}}]_{1:3}$. Based on \eqref{eq:pdf_X_pos}, we can immediately obtain the \acp{PDF} of the positioning error vectors $\mathbf e_{n\mathrm{D}}$ for any $\mathcal{L}(\tilde{\mathbf{v}}_1)$ ($n=1$), $\mathcal{P}(\tilde{\mathbf{v}}_1, \tilde{\mathbf{v}}_2)$ ($n=2$), or $\mathbb{R}^3$ ($n=3$), which are all \acp{GM} with the same number of terms and weights as \eqref{eq:pdf_X_pos}. Specifically, for the 3D positioning error vector $\mathbf e_{\mathrm{3D}}$:
\begin{align}\label{eq:pdf:Error:3D}
    p_{\boldsymbol{E}_{\mathrm{3D}}}(\mathbf e_{\mathrm{3D}}) = \sum\nolimits_{l=1}^L  w^{(l)}_{\boldsymbol{X}} \, f_{\boldsymbol{E}_{\mathrm{3D}}}(\mathbf{e}_{\mathrm{3D}};  {\mathbf{m}}^{(l)}_{3\mathrm{D}} , {\mathsf{\Sigma}}\!\,^{(l)}_{3\mathrm{D}} ),
\end{align}
where, for $l=1,\ldots,L$, 
\begin{align}
   \mathbf{m}^{(l)}_{3\mathrm{D}} =  [\mathbf{m}^{(l)}_{\boldsymbol{X}}]_{1:3} -  \hat{\mathbf{x}}_\mathrm{u}, \quad
   \mathsf{\Sigma}\!\,^{(l)}_{3\mathrm{D}} = [\mathsf{\Sigma}\!\,^{(l)}_{\boldsymbol{X}}]_{1:3,1:3}. 
\end{align}
Letting $\widetilde{\mathsf V}_{1\mathrm{D}}=[\tilde{\mathbf{v}}_1 ]$ and $\widetilde{\mathsf V}_{2\mathrm{D}}=[\tilde{\mathbf{v}}_1 \ \tilde{\mathbf{v}}_2]$, the \acp{PDF} of $\mathbf e_{1\mathrm{D}}$ and $\mathbf e_{2\mathrm{D}}$ are given by
\begin{align}\label{eq:pdf:Error:1D2D}
    p_{\boldsymbol{E}_{n\mathrm{D}}}(\mathbf e_{n\mathrm{D}}) = \sum\nolimits_{l=1}^{L}  w^{(l)}_{\boldsymbol{X}} \, f_{\boldsymbol{E}_{n\mathrm{D}}}(\mathbf e_{n\mathrm{D}}; \mathbf m^{(l)}_{n\mathrm{D}}, \mathsf\Sigma^{(l)}_{n\mathrm{D}}),
\end{align}
where $n=1,2$, and for $l=1,\ldots,L$, 
\begin{align}
 \mathbf m^{(l)}_{n\mathrm{D}} = \widetilde{\mathsf V}_{n\mathrm{D}}^\T \mathbf{m}^{(l)}_{3\mathrm{D}}, 
 \quad  
 \mathsf\Sigma^{(l)}_{n\mathrm{D}} = \widetilde{\mathsf V}_{n\mathrm{D}}^\T {\mathsf{\Sigma}}\!\,^{(l)}_{3\mathrm{D}} \widetilde{\mathsf V}_{n\mathrm{D}}.
\end{align}

To determine the exact 1D \ac{PL} for the subspace $\mathcal{L}(\tilde{\mathbf{v}}_1)$, we use \eqref{eq:pdf:Error:1D2D}. Given that $\mathbf e_{1\mathrm{D}}$, $\mathbf m^{(l)}_{1\mathrm{D}}$, and $\mathsf\Sigma^{(l)}_{1\mathrm{D}}$ are scalars, we denote them as $e_{1\mathrm{D}}$, $m^{(l)}_{1\mathrm{D}}$, and $[\sigma^{(l)}_{1\mathrm{D}}]^2$, respectively. Using the $Q$ function, the actual IR associated with $\hat{\mathbf{x}}_{\mathrm u}$ and any $r$ in $\mathcal{L}(\tilde{\mathbf{v}}_1)$ is expressed as:
\begin{align}
    &\prob{|e_{\mathrm{1D}}|>r }
    = \prob{e_{\mathrm{1D}} < -r } + \prob{e_{\mathrm{1D}} >r } \nonumber\\
    =\,& \sum_{l=1}^{L} w^{(l)}_{\boldsymbol{X}} \left[ 1 - 
    Q \Bigg( \frac{-r \!-\! m^{(l)}_{1\mathrm{D}} }{\sigma^{(l)}_{1\mathrm{D}} }\Bigg) + Q \Bigg( \frac{r\! -\! m^{(l)}_{1\mathrm{D}} }{\sigma^{(l)}_{1\mathrm{D}} }\Bigg)\right] \label{eq:actual:IR:1D}
\end{align}
The smallest $r$ ensuring $\prob{|e_{\mathrm{1D}}|>r } < P_{\mathrm{TIR}}$ is found using bisection search, yielding $\mathrm{PL}_{1\mathrm{D}}$. This value represents the most stringent 1D PL for the position estimate $\hat{\mathbf{x}}_{\mathrm u}$.

To obtain overestimates for the 2D \ac{PL} for $\mathcal{P}(\tilde{\mathbf{v}}_1, \tilde{\mathbf{v}}_2)$ and the 3D PL for $\mathbb{R}^3$ using Lemma~\ref{lemma1}, we select $w_i = 1/n$ for $i=1,...,n$, and find the smallest $r_i$ that ensures 
\begin{align*}
 &\sum_{l=1}^{L} w^{(l)}_{\boldsymbol{X}} \left[ 1 - 
    Q \Bigg( \frac{-r_i \!-\! [\mathbf m^{(l)}_{n\mathrm{D}}]_i }{[\mathsf\Sigma^{(l)}_{n\mathrm{D}}]_{i,i}^{1/2} }\Bigg) + Q \Bigg( \frac{r_i\! -\! [\mathbf m^{(l)}_{n\mathrm{D}}]_i }{[\mathsf\Sigma^{(l)}_{n\mathrm{D}}]_{i,i}^{1/2} }\Bigg)\right] \\
    < & w_iP_{\mathrm{TIR}}
\end{align*}
using bisection search and assigned to $\mathrm{PL}_i$. Finally, the overestimate $\mathrm{PL}_{n\mathrm{D}}^U$ is computed following \eqref{eq:PL:overestimate}. 

\vspace{-2mm}
\subsection{Exact 2D/3D PL Computation}
\label{sec:BayesianII:b}

As discussed in Section~\ref{sec:Problem:b}, computing the exact 2D/3D PL equivalent to finding the minimum radius of a 2D circle or 3D sphere centered at the origin that includes the positioning error vector $\mathbf{e}_{n\mathrm{D}}$ with a probability of at least $1 - P_\mathrm{TIR}$. This is a complex problem. In Appendix~\ref{App:Ellip:PR}, we derive Theorem~\ref{theorem1}, which provides a formulation for the minimum radius and forms the basis for our PL searching algorithm. 

\begin{theorem}\label{theorem1}
Given that the $n$-dimensional positioning error vector $\mathbf{e}_{n\mathrm{D}}$ follows a \ac{GM} distribution with the \ac{PDF}  \eqref{eq:pdf:Error:3D} for $n=3$ and \eqref{eq:pdf:Error:1D2D} for $n=2$, the PL computation problem defined by \eqref{eq:PL:ForAll} can be reformulated as:
\begin{align}\label{eq:PL:2D3D}
\mathrm{PL}_{n\mathrm{D}} = \min\Big\{ r \,\Big|  \sum_{l=1}^{L}  w^{(l)}_{\boldsymbol{X}} \big[1- F_{Z_l}(r^2)\big] < P_{\mathrm{TIR}} \Big\},
\end{align}
where $F_{Z_l}(z) \triangleq \prob{Z_l\leq z}$, $l=1,...,L$, represents the \ac{CDF} of a random variable $Z_l$, given by a weighted sum of $n$ independent noncentral chi-square distributed random variables $W_{l,1},...,W_{l,n}$, each with one degree of freedom and noncentrality parameter $\nu_{l,i}^2$. Specifically, 
\begin{align}\label{eq:Yl}
    Z_l =  \sum_{i=1}^{n} \omega_{l,i} W_{l,i}, \quad W_{l,i} \sim \chi^2\big(1, \nu_{l,i}^2\big), \, \forall i=1,...,n; 
\end{align}
and $\{\omega_{l,i}\}$ and $\{\nu_{l,i}^2\}$ are determined by $\mathsf\Sigma^{(l)}_{n\mathrm{D}}$ and $\mathbf m^{(l)}_{n\mathrm{D}}$ in the following way: Perform eigendecomposition to obtain $\mathsf\Sigma^{(l)}_{n\mathrm{D}} = \mathsf{P}_l \mathsf{\Omega}_l \mathsf{P}_l^\T$, where $\mathsf{P}_l \in \mathbb{R}^{n\times n}$ is an orthogonal matrix and $\mathsf{\Omega}_l$ is a diagonal matrix. The diagonal elements of $\mathsf{\Omega}_l$ are 
$\omega_{l,1},\dots,\omega_{l,n}$ (since $\mathsf\Sigma^{(l)}_{n\mathrm{D}}\succ 0$, $\omega_{l,i}>0$ for $i=1,..., n$); and the noncentrality parameters $\nu_{l,1},...,\nu_{l,n}$ are computed by $(\nu_{l,1},...,\nu_{l,n})^\T \triangleq \mathsf{P}_l^\T [\mathsf\Sigma^{(l)}_{n\mathrm{D}}]^{-\frac{1}{2}} \mathbf m^{(l)}_{n\mathrm{D}}$. 
\begin{proof} 
See Appendix~\ref{App:Ellip:PR}. 
\end{proof}
\end{theorem}

There is no closed-form expression for the \ac{CDF} of a generalized chi-squared variable, but numerical methods are available. In Appendix~\ref{App:Imholf}, we describe the Imhof method \cite{bodenham2016comparison}, a numerical method that can achieve arbitrary accuracy. The CDF $F_{Z_l}(z)$ can be approximated by:
\begin{align}\label{eq:CDF:Yl:approx} 
    \bar{F}_{Z_l}(z, U_l) = \frac{1}{2} - \frac{1}{\pi} \int_0^{U_l} \frac{\sin \beta(u,z)}{u \kappa(u)} \mathrm{d} u ,
\end{align}
where $\beta(u,z)$ and $\kappa(u)$ are given by \eqref{eq:CDF_imhof2} and \eqref{eq:CDF_imhof3}. The approximation error is bounded by 
\begin{align}
    \left|F_{Z_l}(z) - \bar{F}_{Z_l}(z, U_l) \right| \leq \Xi(U_l)
\end{align}
where $\Xi(\cdot)$ is a decreasing function provided in \eqref{eq:TU}. For any desired accuracy, a sufficiently large $U_l$ can be found to ensure $\Xi(U_l)$ meets the requirements. For PL computation, the smallest $U_l$ satisfying $\Xi(U_l) \leq \zeta_1 P_{\mathrm{TIR}}$, where $\zeta_1 \ll 1$, is chosen. To incorporate the approximation error, the constraint in \eqref{eq:PL:2D3D} is modified to:
\begin{align}\label{eq:conndition1_1}
    \sum\nolimits_{l=1}^{L}  w^{(l)}_{\boldsymbol{X}} \big[1- \bar{F}_{Z_l}(r^2,U_l) \big] < (1-\zeta_1 )P_{\mathrm{TIR}}. 
\end{align}

With this numerical tool, $\mathrm{PL}_{n\mathrm{D}}$ in \eqref{eq:PL:2D3D} can be obtained using a search process similar to the bisection search in Section~\ref{sec:Baseline:d}. In particular, the overestimate $\mathrm{PL}_{n\mathrm{D}}^U$ serves as a good initial value for $R_{\mathrm{up}}$. Choosing an initial value for $R_{\mathrm{low}}$ is trivial. During each iteration, the numerical integral \eqref{eq:CDF:Yl:approx} needs to be computed $L$ times, where $L = 2^{M}$. This can be computationally expensive. To reduce the computational complexity, we sort $\{w^{(l)}_{\boldsymbol{X}}\}$ in non-increasing order, denoted by $\{w^{(l(1))}_{\boldsymbol{X}}, w^{(l(2))}_{\boldsymbol{X}}, ..., w^{(l(L))}_{\boldsymbol{X}}\}$. Then we find the smallest integer $J$ such that 
\begin{align}\label{eq:constraintL}
    \sum\nolimits_{j=J+1}^{L} w^{(l(j))}_{\boldsymbol{X}} \leq \zeta_2 \,P_\mathrm{TIR},
\end{align}
where $\zeta_2 \ll 1$. Since $\sum_{l=1}^{L} w^{(l)}_{\boldsymbol{X}} \big[1- F_{Z_l}(r^2)\big] <\sum_{j=1}^{J}  w^{(l(j))}_{\boldsymbol{X}} \big[1-F_{Z_{l(j)}}(r^2)\big] + \sum_{j=J+1}^{L}\! w^{(l(j))}_{\boldsymbol{X}}$, we can replace the constraint in \eqref{eq:PL:2D3D} with a stronger one: 
\begin{align}
    \sum\nolimits_{j=1}^{J}  w^{(l(j))}_{\boldsymbol{X}} \big[1- F_{Z_{l(j)}}(r^2)\big] < (1-\zeta_2) P_\mathrm{TIR}.
\end{align}
When using \eqref{eq:CDF:Yl:approx} for approximation and choosing $U_l$  as described,  \eqref{eq:conndition1_1} becomes
\begin{align*}
   \sum\nolimits_{j=1}^{J}  w^{(l(j))}_{\boldsymbol{X}} \big[1- \bar{F}_{Z_{l(j)}}(r^2, U_{l(j)})\big] < (1\!-\!\zeta_1\!-\!\zeta_2) P_\mathrm{TIR}. 
\end{align*}

The PL search algorithm developed based on the above discussions is summarized in Algorithm~1. Due to conservative approximations, the output is expected to be looser than the optimal PL. The gap depends on the choice of $\zeta_1$ and $\zeta_2$, the larger they are, the looser the computed PL.

\begin{algorithm}[t]
\label{alg:1}
\caption{Exact $n$-dimensional PL computation, $n = 2, 3$}
\begin{algorithmic}[1] 
\Require $P_\mathrm{TIR}$, $p_{\boldsymbol{E}_{n\mathrm{D}}}(\mathbf e_{n\mathrm{D}})$ given by \eqref{eq:pdf:Error:3D} or \eqref{eq:pdf:Error:1D2D}, $\zeta_1$, $\zeta_2$, search error tolerance $r_\mathrm{tol}$.
\Ensure $\mathrm{PL}_{n\mathrm{D}}$ defined in \eqref{eq:PL:2D3D} within error tolerance $r_\mathrm{tol}$. 
\State Compute the overestimate $\mathrm{PL}_{n\mathrm{D}}^U$ following Section~\ref{sec:BayesianII:a};
\State Sort $\{w^{(l)}_{\boldsymbol{X}}\}$ into non-increasing order and find the smallest $J$ satisfying \eqref{eq:constraintL};
\For{$j = 1,...,J$} \Comment{Parameter preparation}
\State Compute  $\{\omega_{l(j),i}\}$ and $\{\nu_{l(j),i}^2\}$ following Theorem~\ref{theorem1}; 
\State Find $U_{l(j)}$ such that $\Xi(U_{l(j)}) = \zeta_1 P_{\mathrm{TIR}}$ based on \eqref{eq:TU};
\EndFor
\State $r_{\mathrm{up}} \gets \mathrm{PL}_{n\mathrm{D}}^U$; \Comment{Starting upper limit for $\mathrm{PL}_{n\mathrm{D}}$}
\State $r_{\mathrm{low}} \gets$ some small value; \Comment{Starting lower limit} 
\While{$|r_{\mathrm{up}} -r_{\mathrm{low}}| > r_\mathrm{tol}$}
\State $r_{\mathrm{mid}} \gets (r_{\mathrm{up}} + r_{\mathrm{low}})/2$;
\State $P_{\mathrm{mid}} \gets  \sum_{j=1}^{J}  w^{(l(j))}_{\boldsymbol{X}} \big[1- \bar{F}_{Z_{l(j)}}(r_{\mathrm{mid}}^2, U_{l(j)} )\big]$ using  \eqref{eq:CDF:Yl:approx};
\If{$P_{\mathrm{mid}} < (1-\zeta_1-\zeta_2) P_\mathrm{TIR}$}
    \State $r_{\mathrm{up}} \gets r_{\mathrm{mid}}$, 
    $P_{\mathrm{up}} \gets P_{\mathrm{mid}}$;
\Else
    \State $r_{\mathrm{low}} \gets r_{\mathrm{mid}}$,
    $P_{\mathrm{low}} \gets P_{\mathrm{mid}}$;
\EndIf
\EndWhile
\State $\mathrm{PL}_{n\mathrm{D}} \gets r_{\mathrm{up}}$. 
\end{algorithmic}

\end{algorithm}

\begin{remark}[Complexity discussion]
\label{remark:complexity:bayesianII}
Computing the exact 1D PL using bisection search is efficient, as is the process of overestimating the 2D/3D PL. For a 1D search with $N_\mathrm{it}>0$ iterations, the computational complexity is $\mathcal{O}(2 N_\mathrm{it} L)$. In contrast, obtaining the exact 2D/3D PL using Algorithm~1 is more computationally intensive due to the numerical integration required in \eqref{eq:CDF:Yl:approx}. To ensure accuracy, the upper integration limit $U_l$ is typically set high by choosing a small $\zeta_1 \ll 1$. Our numerical studies show that the integration process is also sensitive to the parameters of the generalized chi-squared variables in \eqref{eq:Yl}, which are dependent on the measurement bias distributions. Additionally, discarding GM terms with negligible weights, as described in \eqref{eq:constraintL}, can significantly reduce the computational complexity across all methods. However, increasing the number of discarded terms (i.e., using a larger $\zeta_2$) can cause conservative PL values. Therefore, $\zeta_1$ and $\zeta_2$ should be carefully selected to balance computational complexity and the tightness of the PL. 
\end{remark}

\vspace{-3mm}
\section{Baseline RAIM Algorithm}
\label{sec:Baseline}

To benchmark performance, we adapt the advanced \ac{RAIM} algorithm from \cite{blanch2015baseline}. We refer to the adapted algorithm as the \textit{Baseline RAIM}\footnote{The dropped term ``advanced'' emphasizes its ability to detect and exclude multiple faults, which is more advanced than the earliest \ac{RAIM} algorithm (which could handle only one fault).} algorithm. It uses a hypothesis testing framework for \ac{FDE}. The following inputs are required: $P_\mathrm{TIR,H}$ and $P_\mathrm{TIR,V}$, the \ac{TIR} requirements for the horizontal-plane and the vertical direction, and $P_\mathrm{FA, H}$ and $P_\mathrm{FA,V}$, the respective false alarm budgets. A false alarm occurs if the algorithm detects faults in fault-free measurements. The outputs include a position estimate and PLs for both horizontal and vertical subspaces. 
The main steps of Baseline RAIM are as follows. 
\begin{enumerate}
    \item \textbf{Fault mode determination}: Identify a set of \emph{fault modes} (hypotheses) for monitoring, where each fault mode assumes a subset of measurements is faulty and the rest are fault-free. These fault modes are exclusive, meaning only one can occur.
    \item \textbf{SS test for fault detection}: Compute an SS test statistic for each fault mode and position coordinate to evaluate the closeness of different position estimates obtained from assumed fault-free measurements by the fault modes.
    \item \textbf{Fault exclusion (if SS test fails)}: If the SS test fails, attempt fault exclusion by repeatedly applying fault detection to subsets of assumed fault-free measurements for each fault mode, starting with the one with the highest probability of occurrence (see Remark~\ref{remark1}).
    \item \textbf{Position estimate and PL calculation}: If fault exclusion is successful or not needed, calculate a position estimate and compute two PLs following \cite{blanch2015baseline}: one for the horizontal plane and one for the vertical direction.
\end{enumerate}

In this process, the following components are involved: (i) all-in-view position estimation;  (ii) fault mode identification; (iii) \ac{SS} testing; and (iv) \ac{PL} computations. The following subsections provide detailed explanations of each component, and conclude with a discussion of computational complexity.

\vspace{-3mm}
\subsection{All-in-View Position Estimation}
\label{sec:Baseline:a}

The all-in-view position assumes all measurements are fault-free. In the linear model \eqref{eq:linearized-3D-model:matrix}, when $\mathbf{b} = \mathbf{0}$ and $\mathbf n$ follows $\mathcal N(\mathbf{0}, \mathsf{\Sigma})$, where $\mathsf{\Sigma}$ is the diagonal covariance matrix with elements $\mathsf{\Sigma}_{i,i} = \sigma_{n,i}^2$, $i=1,\ldots,M$, the \ac{WLS} estimate of $\mathbf{x}$, also the \ac{ML} estimate under under the fault-free assumption, is given by 
\begin{align}\label{eq:Xhat:0}
    \hat{\mathbf{x}}^{(0)} = ( \mathsf{H}^\T \mathsf{\Sigma}^{-1} \mathsf{H} )^{-1} \mathsf{H}^\T \mathsf{\Sigma}^{-1} \mathbf y = \mathsf{A}^{(0)} \mathbf y, 
\end{align}
where $\mathsf{A}^{(0)} \triangleq ( \mathsf{H}^\T \mathsf{\Sigma}^{-1} \mathsf{H} )^{-1} \mathsf{H}^\T \mathsf{\Sigma}^{-1} $. 
The positioning error vector $\mathbf{x} -\hat{\mathbf{x}}^{(0)} = \mathsf{A}^{(0)} \mathbf n$ follows a zero-mean Gaussian distribution with covariance matrix $\mathsf{\Phi}^{(0)} \triangleq ( \mathsf{H}^\T \mathsf{\Sigma}^{-1} \mathsf{H} )^{-1}$. 

\vspace{-3mm}
\subsection{Fault Modes Identification}
\label{sec:Baseline:b}

To enable \ac{SS} testing for fault exclusion, an eligible fault mode must contain at least $5$ assumed fault-free measurements. Baseline RAIM monitors all eligible modes for optimal performance. The total number of monitored fault modes (with at least one fault) is given by $N_{\mathrm{FM}} = \sum_{j=1}^{M-5} \binom{M}{j}$. 
For fault mode $k = 1, ..., N_{\mathrm{FM}}$, let $\mathcal{I}_k$ represent the index set of assumed fault-free measurements, and  $\mathcal{I}_k^c \triangleq \{1,2,\ldots,M\} \setminus \mathcal{I}_k$ represent the index set of faulty  measurements. The probability of occurrence of fault mode $k$ is $p_{\mathrm{FM},k} = \prod\nolimits_{i \in \mathcal{I}_k} (1 - \theta_i) \prod\nolimits_{i \in \mathcal{I}_k^c} \theta_i$. 
For convenience, we assume that fault modes are sorted in decreasing order of probability of occurrence, thus $p_{\mathrm{FM},k} \geq p_{\mathrm{FM},k+1}$, for $1\leq k<N_{\mathrm{FM}}$. 
\vspace{-3mm}

\subsection{Solution Separation Testing}
\label{sec:Baseline:c}

For fault mode $k$, define $\mathsf{\Sigma}_k$ as the diagonal matrix derived from $\mathsf{\Sigma}$ in the following way
\begin{align}
  [\mathsf{\Sigma}_k^{-1}]_{i,i} = \begin{cases}
        0, & i\in \mathcal I_k^c, \\
        1/\sigma_{n,i}^{2}, & i \in  \mathcal I_k.
    \end{cases} 
\end{align}
The \ac{WLS} estimate of $\mathbf{x}$ using measurements indexed by $\mathcal{I}_k$ is 
\begin{align}\label{eq:x_hat_k}
    \hat{ \mathbf{x}}^{(k)} =\mathsf{A}^{(k)} \mathbf{y} 
\end{align}
where $\mathsf{A}^{(k)} \triangleq  \left( \mathsf{H}^\T \mathsf{\Sigma}_k^{-1} \mathsf{H} \right)^{-1} \mathsf{H}^\T \mathsf{\Sigma}_k^{-1}$. 
If all these measurements are fault-free, the positioning error $\mathbf{x} -\hat{\mathbf{x}}^{(k)} = \mathsf{A}^{(k)} \mathbf n$ follows a zero-mean Gaussian distribution with covariance matrix $\mathsf{\Phi}^{(k)} \triangleq ( \mathsf{H}^\T \mathsf{\Sigma}_k^{-1} \mathsf{H} )^{-1}$, and difference between the all-in-view and fault mode $k$ estimates
\begin{align}
     \Delta \hat{\mathbf{x}}^{(k)} \triangleq \hat{\mathbf{x}}^{(0)} -\hat{\mathbf{x}}^{(k)}
\end{align}
follows a zero-mean Gaussian distribution with  covariance matrix given by $( \mathsf{A}^{(k)} -\mathsf{A}^{(0)}) \mathsf{\Sigma} ( \mathsf{A}^{(k)} -\mathsf{A}^{(0)})^{\mathrm T}$. The \ac{SS} test statistics used by the Baseline RAIM algorithm are 
\begin{align}\label{eq:tau:nk}
    \tau_{n,k} \triangleq \big|\Delta \hat{\mathbf{x}}^{(k)}_n \big|, 
    \quad n = 1,2,3,\  k = 1, \dots, N_{\mathrm{FM}}.
\end{align}
 
The \ac{SS} test compares $\tau_{n,k}$ with a test threshold $T_{n,k}$ for each position coordinate and fault mode. Only if $\tau_{n,k} \leq T_{n,k}$ for all $n = 1,2,3$ and $k = 1, \dots, N_{\mathrm{FM}}$, are the $M$ measurements considered fault-free, and the algorithm outputs $\hat{\mathbf{x}}_{\mathrm u} = \hat{\mathbf{x}}^{(0)}_{1:3}$ as the 3D position estimate. If $\tau_{n,k} > T_{n,k}$ for any $n,k$, the algorithm proceeds to fault exclusion.  
The test thresholds are determined based on the false alarm budgets $P_\mathrm{FA, H}$ and $P_\mathrm{FA,V}$\footnote{Reducing $P_\mathrm{FA,H}$ and $P_\mathrm{FA,V}$ increases the test thresholds computed by \eqref{eq:T_ss_test}, thus lowering the likelihood of issuing warning for potential faults. This reduces computational costs but raises the risk of misdetection. Conversely, higher $P_\mathrm{FA,H}$ and $P_\mathrm{FA,V}$ trigger more fault exclusion attempts, raising computational costs but enhancing positioning quality.}. 
The algorithm distributes $P_\mathrm{FA, H}$ evenly between the two horizontal position coordinates $x_1$ and $x_2$, and then distributes each coordinate's false alarm budget evenly across the $N_{\mathrm{FM}} $ fault modes. For $k=1,\ldots, N_{\mathrm{FM}}$, the thresholds are
\begin{align}\label{eq:T_ss_test}
    T_{n,k} = \begin{cases}
        \sigma_{ss,n}^{(k)} \, Q^{-1} \left(\frac{P_\mathrm{FA,H}}{4 N_{\mathrm{FM}} }\right),  & n = 1,2 \\
        \sigma_{ss,n}^{(k)} \, Q^{-1} \left(\frac{P_\mathrm{FA,V}}{2 N_{\mathrm{FM}} }\right),  & n = 3
    \end{cases}
\end{align}
where  $\sigma_{ss,n}^{(k)} \triangleq  
    \big[ ( \mathsf{A}^{(k)} -\mathsf{A}^{(0)}) \mathsf{\Sigma} ( \mathsf{A}^{(k)} -\mathsf{A}^{(0)})^{\mathrm T} \big]_{n,n}^{1/2}$, and $Q^{-1}( \cdot)$ is the inverse of the $Q$ function:  $Q(u) = \frac{1}{\sqrt{2\pi}}\int_u^{+\infty} e^{-\frac{t^2}{2}} \mathrm d t$.

\begin{remark}[Fault exclusion process]
\label{remark1}
When the \ac{SS} test fails, the algorithm attempts to exclude the faulty measurements by reapplying the fault mode identification and SS testing to subsets of assumed fault-free measurements specified by the $N_\mathrm{FM}$ fault modes, ordered by decreasing probability of occurrence. This continues until the SS test passes or all fault modes are checked, indicating fault exclusion failure. Upon reaching fault mode $k$ in this process, the subset of measurements indexed by $\mathcal{I}_k$ are checked, and a new list of $N_{\mathrm{FM},k} = \sum_{j=1}^{|\mathcal I_k| - 5} \binom{|\mathcal I_k|}{j}$ fault modes is identified. If the \ac{SS} testing is successful, fault mode $k$ is considered true, and the algorithm returns the position estimate ($\hat{\mathbf{x}}_{\mathrm u} = \hat{\mathbf{x}}^{(k)}_{1:3}$) and \acp{PL} computed  using these $|\mathcal I_k|$ measurements. If fault exclusion fails, the algorithm declares that the system cannot provide reliable positioning results (system unavailable).
\end{remark}

\vspace{-4mm}
\subsection{Protection Level Computation}
\label{sec:Baseline:d}

The Baseline RAIM algorithm can compute 1D \acp{PL} for each of the three coordinate directions. It can also obtain PL overestimates in the three coordinate planes and the full 3D space using Lemma~\ref{lemma1}. Below we summarize the formulation of the 1D PL for the vertical direction and the 2D PL overestimate for the horizontal plane as described in \cite{blanch2015baseline}.  Assume that the SS test passes for the $M$ measurements, so $\hat{\mathbf{x}}_{\mathrm u} = \hat{\mathbf{x}}^{(0)}_{1:3}$, and we define $\sigma_n^{(k)} \triangleq [ \mathsf{\Phi}^{(k)} ]_{n,n}^{1/2}$ for $n=1,2,3$ and $k = 0,1,\dots, N_{\mathrm{FM}}$.

In the vertical direction, the actual \ac{IR} associated with an arbitrary $r$ is given by $\mathrm{Pr} \{|x_3 - \hat{x}^{(0)}_3 |> r\}$. An overbound of the vertical PL $\mathrm{PL}_{\mathrm V}$ is given by the minimum $r$ satisfying the following condition
\begin{align}\label{eq:VPL_ARAIM}
   2  Q \left( \frac{ r }{\sigma_3^{(0)}} \right) 
    +  \sum_{k=1}^{N_\mathrm{FM}} p_{\mathrm{FM},k} \, Q\!\left( \frac{r\!-\! T_{3,k}}{\sigma_3^{(k)} } \right) < P_{\mathrm{TIR},\mathrm V}. 
\end{align}
A bisection search is used to find the smallest $r$  within a given error tolerance $r_{\mathrm{tol}}$. The detailed formulation process can be found in \cite[Appendix H]{blanch2015baseline}. 
In the horizontal plane, $P_{\mathrm{TIR},\mathrm H}$ is evenly divided for the $\mathrm{x}$ and $\mathrm{y}$ directions (i.e., $w_1=w_2=0.5$ in \eqref{eq:IR:split}). For $n=1,2$, an overbound of the PL $\mathrm{PL}_{n}$ is given by the minimum $r_n$ that satisfies
\begin{align}\label{eq:HPL_ARAIM}
    2 Q \left( \frac{ r_n }{\sigma_n^{(0)}} \right) 
    \!+  \sum_{k=1}^{N_\mathrm{FM}} p_{\mathrm{FM},k} \, Q\!\left( \frac{r_n\,-\,T_{n,k}}{\sigma_n^{(k)} } \right) < \frac{P_{\mathrm{TIR},\mathrm H}}{2}.
\end{align}
Again, bisection search is used to find these minimum values within error tolerance $r_{\mathrm{tol}}$. Following  \eqref{eq:PL:overestimate}, the horizontal PL overestimate is given by $\mathrm{PL}_{\mathrm H} = (\mathrm{PL}_{1}^2 + \mathrm{PL}_{2}^2)^{1/2}$.

If the \ac{SS} test fails for the $M$ measurements and the $k$th fault mode is accepted during the fault exclusion process, then $\hat{\mathbf{x}}_{\mathrm{u}} = \hat{\mathbf{x}}^{(k)}_{1:3}$, and in \eqref{eq:VPL_ARAIM} and \eqref{eq:HPL_ARAIM}, $p_{\mathrm{FM},k}$, $T_{n,k}$, and $\sigma_n^{(k)}$ should be replaced with their counterparts computed for the $N_{\mathrm{FM},k}$ \emph{new} fault modes. 
The left-hand sides of \eqref{eq:VPL_ARAIM} and \eqref{eq:HPL_ARAIM} are loose upper bounds of the actual \ac{IR} related to the 1D positioning error variables \cite{blanch2015baseline}. Moreover, $\mathrm{PL}_{\mathrm H}$, by formulation, is an overestimate of the 2D PL for the horizontal plane. Due to these, the values obtained for $\mathrm{PL}_{\mathrm V}$ and $\mathrm{PL}_{\mathrm H}$ tend to be large. This intrinsic drawback will be addressed by the proposed Bayesian RAIM algorithm.

\begin{remark}[Complexity discussion]
\label{remark:complexity:baseline}
The Baseline RAIM algorithm has a variable computational complexity due to the varying number of SS tests required (from $1$ in the best case and $N_{\mathrm{FM}}$ in the worst case). Most of the computational cost for SS tests comes from the matrix operations in \eqref{eq:Xhat:0}--\eqref{eq:T_ss_test}, with matrix sizes depending on the number of assumed fault-free measurements, ranging between $M$ and $5$. To reduce complexity, fault modes with occurrence probabilities significantly lower than the \ac{TIR} can be excluded from monitoring\cite{blanch2015baseline}.
PL computation via bisection search adds little to the total cost, as confirmed by our numerical study. This iterative algorithm requires a variable number of iterations, denoted by $N_\mathrm{it}>0$, and its complexity is $\mathcal{O}(N_\mathrm{it}N_{\mathrm{FM}})$, since both \eqref{eq:VPL_ARAIM} and \eqref{eq:HPL_ARAIM} involve about $N_{\mathrm{FM}}$ Q-function evaluations per iteration.
\end{remark}

\vspace{-3mm}
\section{Numerical Study}
\label{sec:Numerical:Study}

We conduct numerical studies with two main objectives: (i) to compare the performance and computational complexity (in terms of running time) of the proposed Bayesian RAIM algorithm against the baseline RAIM algorithm; and (ii) to assess how sensitive the measurement model linearization is to the initial positioning error, $\mathbf{e}_0$. 

\vspace{-3mm}
\subsection{Simulation Description}

\subsubsection{Scenario and parameters}

\begin{figure}[!t]
    \centering \vspace{-1.0em}
    \includegraphics[trim={0.5cm .2cm 0.5cm .05cm},clip, width = .65\linewidth]{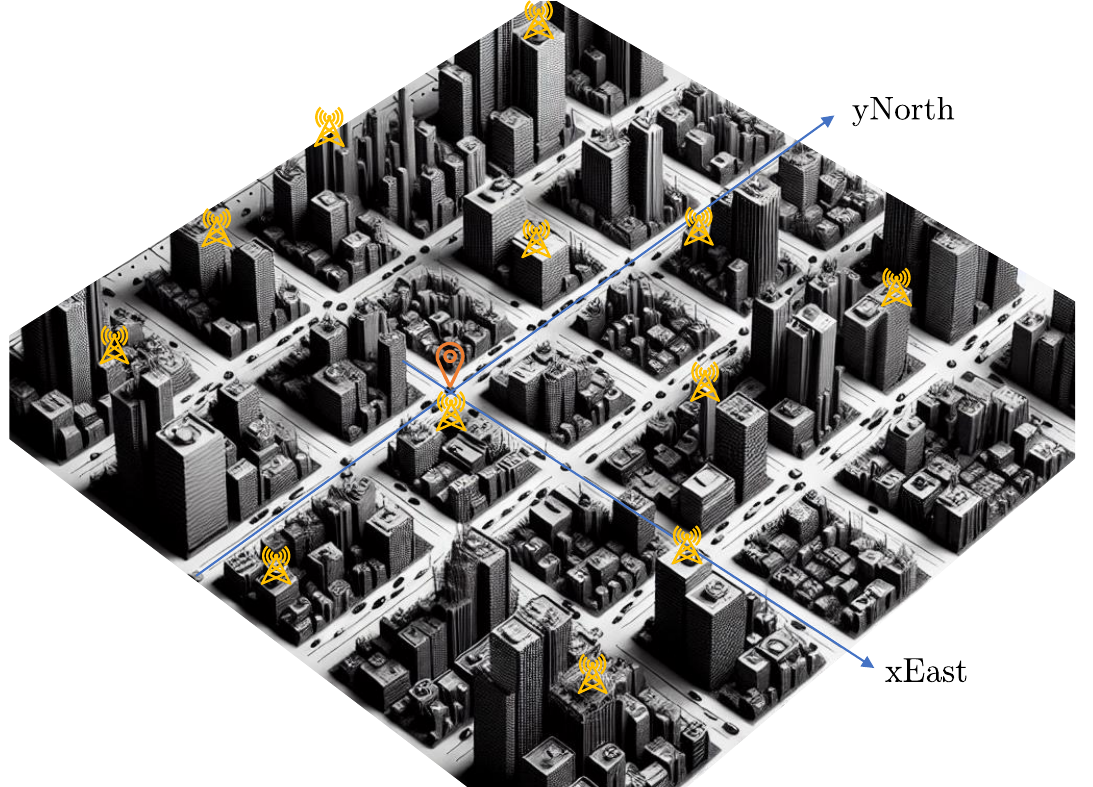}
    \vspace{-.5em}
    \caption{Illustration of the dense urban grid scenario used in the simulation study. BSs are shown in yellow, while the UE location is marked with a red location pin.} 
    \vspace{-1.5em}
    \label{fig:BS:pos}
\end{figure}

The simulation study is conducted in a dense urban grid covering approximately $1000 \times 1000$~m$^2$, divided into $12$ cells of $400 \times 250$~m$^2$ each. A total of $M=12$ \acp{BS} are deployed, one per cell, each placed on a building rooftop at a random height between 10 and 30 meters. The scenario is illustrated in Fig.~\ref{fig:BS:pos}. The position of the target UE is fixed at $\mathbf{x}_{\mathrm u} = (0,0,0)$ with zero clock bias ($x_c =0$). \ac{ToA} measurement noise is identically distributed across all \acp{BS} with zero mean, $\sigma_{n,i} = 0.5$ m \cite[Annex~B2]{3gpp.38.859}, and a prior fault probability $\theta_i = 0.05$. Two fault types are considered: one due to strong \ac{NLoS} signals, with $m_{b,i}$ between 1 and 20 m and $\sigma_{b,i} = 1$ m  for all $i$, and another due to clock synchronization errors, with $m_{b,i} = 0$ m and  $\sigma_{b,i} = 10$ m for all $i$. Note that the NLOS conditions were not generated to match the building locations in Fig.~\ref{fig:BS:pos}. A \ac{TIR} of $P_\mathrm{TIR} = 10^{-3}$ is used for all PL computations. 

\subsubsection{Simulation setup} 

For performance comparison, $N_{\mathrm{sim}} = 5 \times 10^5$ positioning epochs are simulated for each fault type, using an error-free initial \ac{UE} position estimate, $\mathbf{x}_{\mathrm{u},0} = \mathbf{x}_{\mathrm{u}}$. This ensures there is no linearization errors in \eqref{eq:linearization_1}. The Baseline RAIM algorithm computes \acp{PL} for both the horizontal-plane ($\mathrm{H}$-plane) and the vertical-direction ($\mathrm{V}$-direction), with false alarm probabilities set to $P_\mathrm{FA,H} = P_\mathrm{FA,V} = 10^{-2}$. The Bayesian RAIM algorithm computes the horizontal PL overestimate, $\mathrm{PL}_{\mathrm{H}}^U$, using the method described in Section~\ref{sec:BayesianII:a}, a near-optimal $\mathrm{PL}_{\mathrm{H}}$ via the \emph{exact PL computation} algorithm with $\zeta_1 = 0.1$ and $\zeta_2 = 0.002$, and the optimal \acp{PL} for the vertical and for the $45^\circ$ directions in the $\mathrm{x}$-$\mathrm{y}$ plane, where unit vector is $\tilde{\mathbf{v}}_{45^\circ} = (\cos 45^\circ, \sin 45^\circ, 0)^\mathrm{T}$. Additionally, two variations of the Bayesian RAIM algorithm are run for benchmarking. The first, termed the \textit{fault-ignorant variation}, assumes zero fault probability for all measurements, yielding an identical Gaussian posterior PDF of $\boldsymbol{X}$ in all epochs. The PL derived from this PDF  corresponds to the positioning accuracy at the $1-\mathrm{TIR}$ percentile. 
The second, termed the \textit{genie variation}, uses perfect knowledge about the fault state, setting $\theta_i$ to $0$ or $1$ based on the actual fault state. This also results in a Gaussian posterior PDF, but it varies across epochs. Results from this set of simulation are discussed in Section~\ref{sec:performance:part1}. 

Another set of simulations is conducted to evaluate the sensitivity of the two RAIM methods to initial positioning error. An initial error $\mathbf{e}_{0}$ is introduced such that $\mathbf{x}_{\mathrm{u},0} = \mathbf{x}_{\mathrm{u}} + \mathbf{e}_{0}$. Two error models are considered: a horizontal error, $\mathbf{e}_{0,\mathrm{H}} = (E_\mathrm{H} \cos\phi, E_\mathrm{H} \sin\phi, 0)^\T$, where  $\phi$ is uniformly distributed over $[0, 2\pi)$, and a vertical error, $\mathbf{e}_{0,\mathrm{V}} = (0,0,E_\mathrm{V})^\T$. $E_\mathrm{H}$ ranges from $0$ to $5$~m in $0.5$~m increments, while $E_\mathrm{V}$ ranges from $-10$ to $10$~m in $2$~m steps. For each value of $E_\mathrm{H}$ or $E_\mathrm{V}$, $N_{\mathrm{sim}}$ positioning epochs are simulated for each fault type, using identical noise and bias realizations. For Bayesian \ac{RAIM}, we focus on the vertical direction and the horizontal $\tilde{\mathbf{v}}_{45^\circ}$ direction. Results from this set of simulation are discussed in Section~\ref{sec:performance:part2}.

\subsubsection{Performance evaluation metrics}
The algorithms are evaluated using the following metrics: 
\begin{enumerate}
    \item \emph{Simulated \ac{IR}:} Rather than adopting the traditional, requirement-specific definition of \ac{IR}, which depends on both the \ac{PL} threshold and time-to-alert (TTA) constraints, we use a requirement-independent definition of \ac{IR} in this study: 
    \begin{align}
        \mathrm{IR}_{n\mathrm{D}} = \frac{\sum_{k=1}^{N_{\mathrm{sim}}} I(\mathrm{PE}_{n\mathrm{D},k} > \mathrm{PL}_{n\mathrm{D},k}) }{N_{\mathrm{sim}}}, 
    \end{align}
    where $\mathrm{PE}_{n\mathrm{D},k} \triangleq \Vert \mathbf{e}_{n\mathrm{D},k} \Vert$ is the actual positioning error (PE) and $\mathrm{PL}_{n\mathrm{D},k}$ is the computed PL for the $k$th epoch. The indicator function $I(\cdot)$ returns $1$ if $\mathrm{PE}_{n\mathrm{D},k} > \mathrm{PL}_{n\mathrm{D},k}$, and $0$ otherwise. 
    \item \emph{PL tightness:} We assess how closely the \ac{PL} track the actual errors using the \textit{Stanford diagram} \cite{whiton2022cellular}, which plots $\mathrm{PE}_{n\mathrm{D}}$ versus $\mathrm{PL}_{n\mathrm{D}}$ for each epoch on a PE-PL plane\footnote{The resolution and color intensity of a Stanford diagram depends on the density of pixels, each of which contains a small square area.}. Points below the diagonal indicate integrity failures; ideally, points cluster in the lower-left for tight PLs and low errors. PL tightness is further quantified using empirical CDFs, with PL values reported at the $50$th (median), $95$th, and $99$th percentiles.  These percentiles are chosen to represent typical and extreme cases, and the empirical CDF curves in Fig.~\ref{fig:Compare:Genie} illustrate the position of these PL values within the overall distribution.

    \item \textit{Running time:} Computational complexity is measured by the running time in Matlab on a MacBook Pro with an Intel Core i7 processor. For the Bayesian RAIM algorithm, all nine message passing steps described in Section~\ref{sec:BayesianI:a} are performed.
\end{enumerate}

\vspace{-3mm}
\subsection{Performance Comparison without Linearization Error}
\label{sec:performance:part1}

Fig.~\ref{fig:SD:NloS} and Fig.~\ref{fig:SD:Clock} show Stanford diagrams for NLoS- and clock-type faults, respectively, with pixel sizes specified in the captions. Both algorithms achieve simulated IRs below the $P_\mathrm{TIR} = 10^{-3}$ limit in all subspaces, demonstrating their reliability. Bayesian RAIM consistently provides tighter PLs than Baseline RAIM, even though the point cloud shapes vary across subspaces and fault types. Percentile values of the horizontal overestimate $\mathrm{PL}_\mathrm{H}^U$ and the near-optimal $\mathrm{PL}_\mathrm{H}$, along with their reduction over Baseline RAIM, are summarized in Table~\ref{tab:PL}. In the Bayesian RAIM diagrams, two blue dash-dotted lines are shown, one aligned with the $\mathrm{PE}$-axis and one with the $\mathrm{PL}$-axis, which indicate the PL value obtained from the fault-ignorant variation. This value (corresponding to the accuracy at the $99.9\%$ percentile) serves as the theoretical lower bound for the PL computed at any epoch, as demonstrated in our simulation results. The dots appearing to the right of the vertical blue line represent epochs in which the actual PE exceeds this accuracy threshold. The percentage of such occurrences is significantly higher than the simulated \ac{IR}, highlighting the necessity of accounting for potential faults rather than relying solely on nominal accuracy.

\subsubsection{\texorpdfstring{$\mathrm{H}$}{H}-plane} 
\begin{figure*}[!t]
\centering \vspace{-1.5em}
    \subfigure[Bayesian RAIM, ${\mathrm{H}}$-plane Alg.~1]{ \includegraphics[width= .27\linewidth]{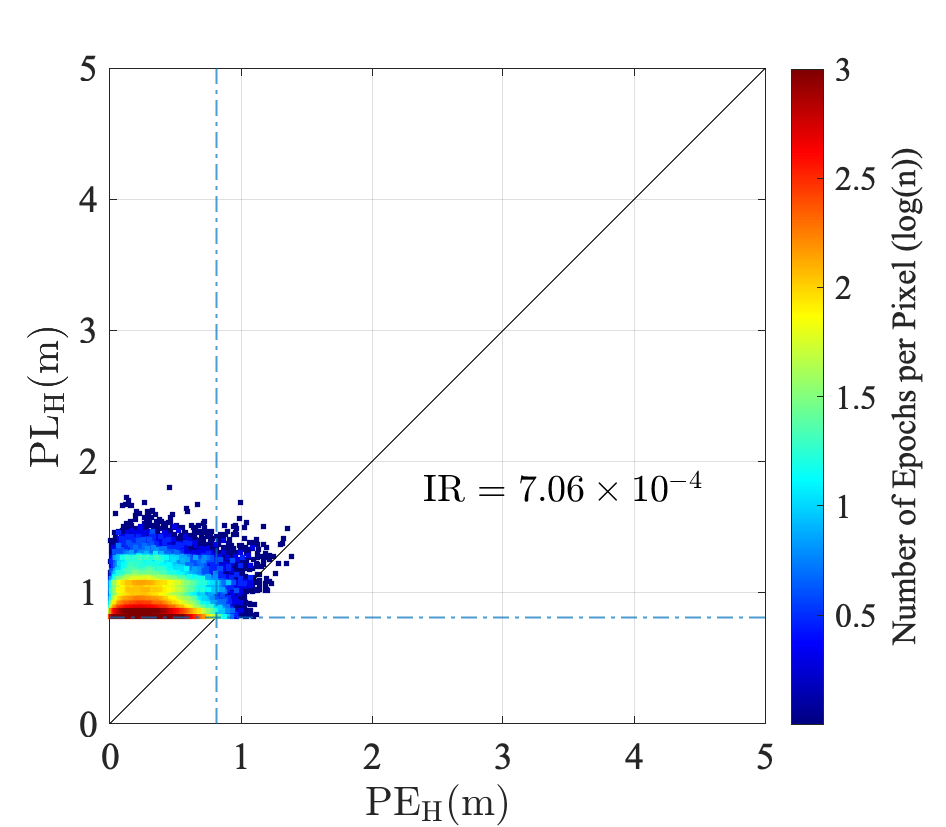}}
 \hfill
    \subfigure[Bayesian RAIM, ${\mathrm{H}}$-plane overestimate]{ \includegraphics[width= .27\linewidth]{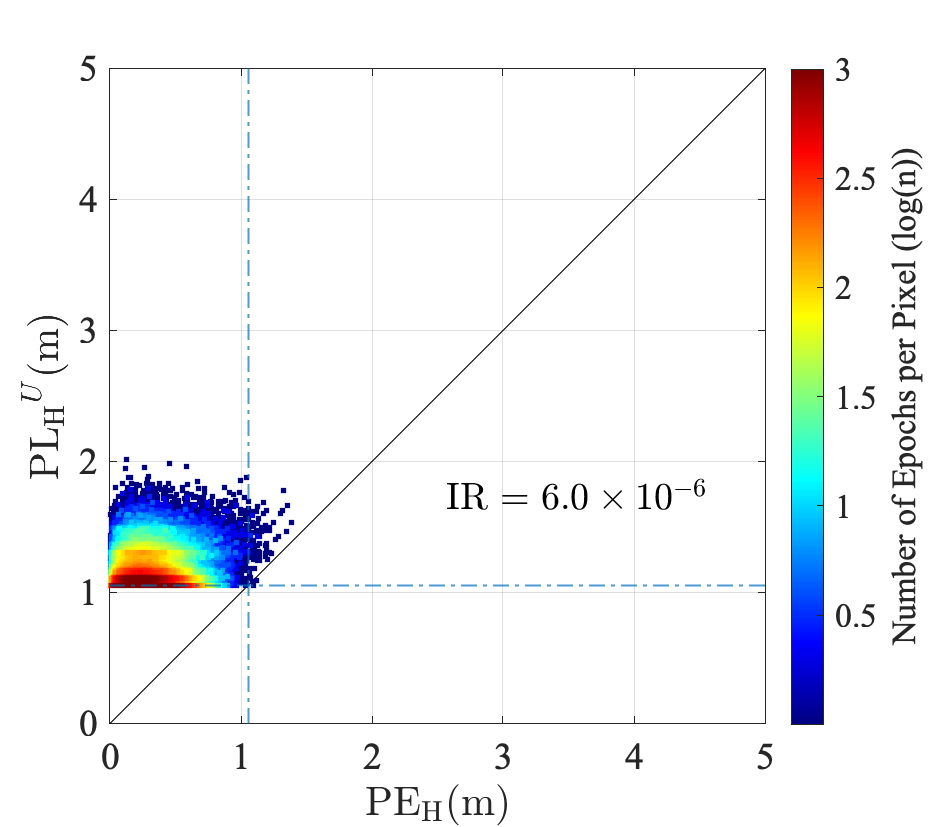}}
 \hfill \vspace{-1em}
 \subfigure[Baseline RAIM, ${\mathrm{H}}$-plane]{ \includegraphics[width= .28\linewidth]{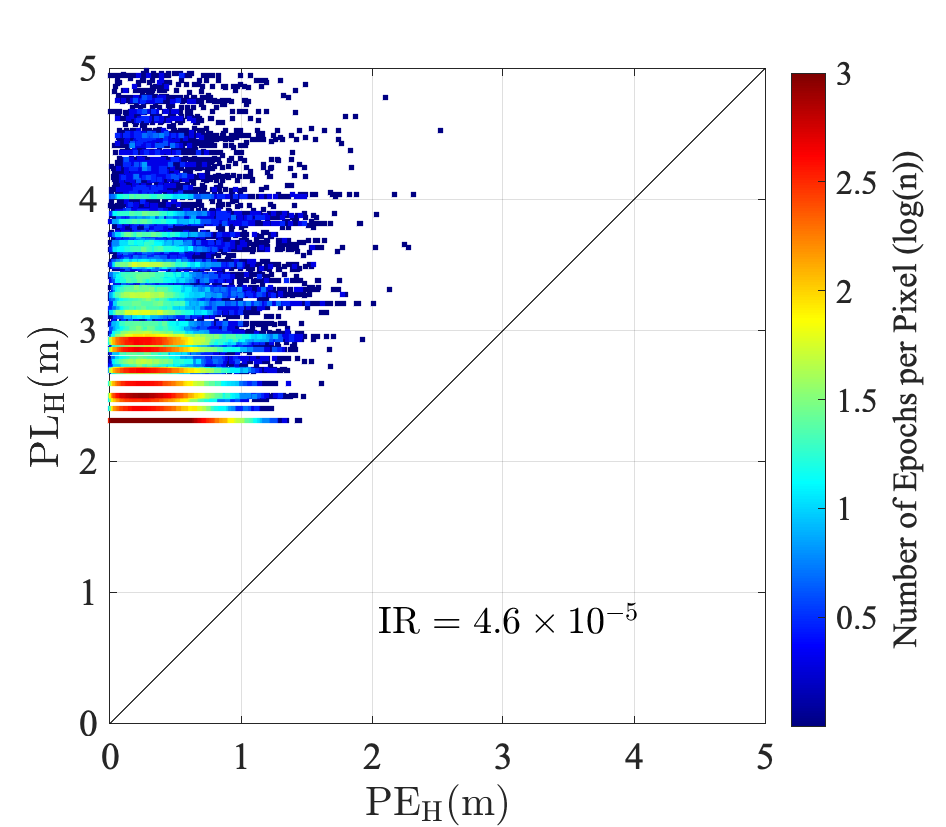}} 
 \subfigure[Bayesian RAIM, $\tilde{\mathbf{v}}_{45^\circ}$-direction]{ \includegraphics[width= .27\linewidth]{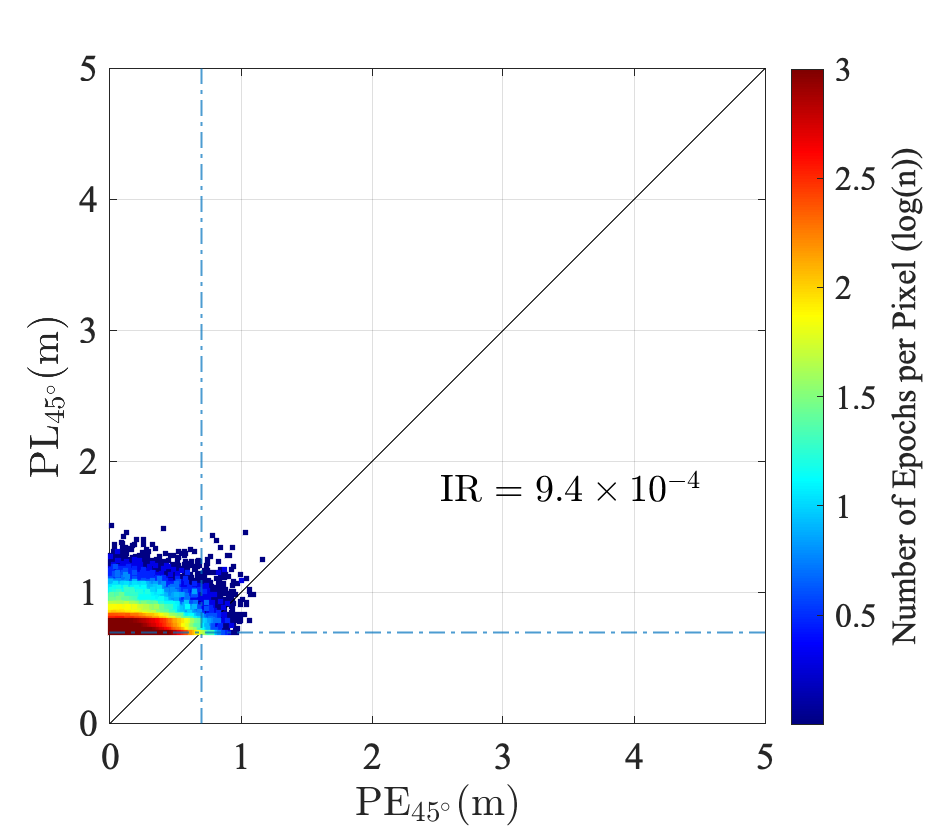}}
 \hfill
    \subfigure[Bayesian RAIM, ${\mathrm{V}}$-direction]{ \includegraphics[width= .27\linewidth]{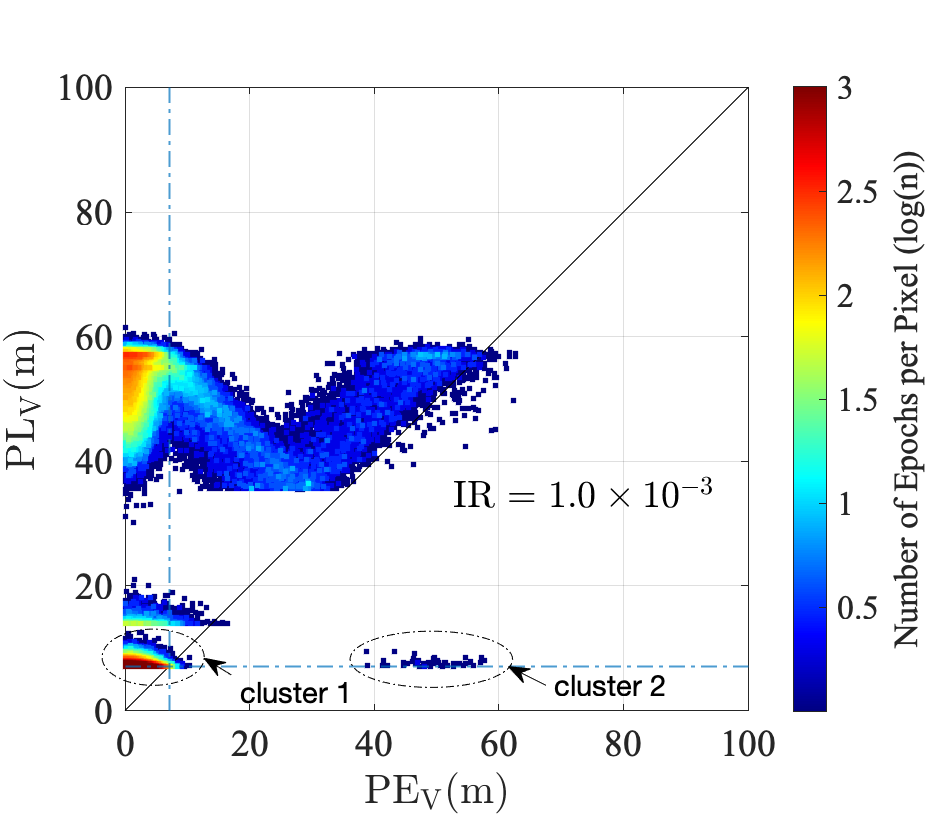}}
  \hfill \vspace{-.5em}
    \subfigure[Baseline RAIM, ${\mathrm{V}}$-direction]{ \includegraphics[width= .27\linewidth]{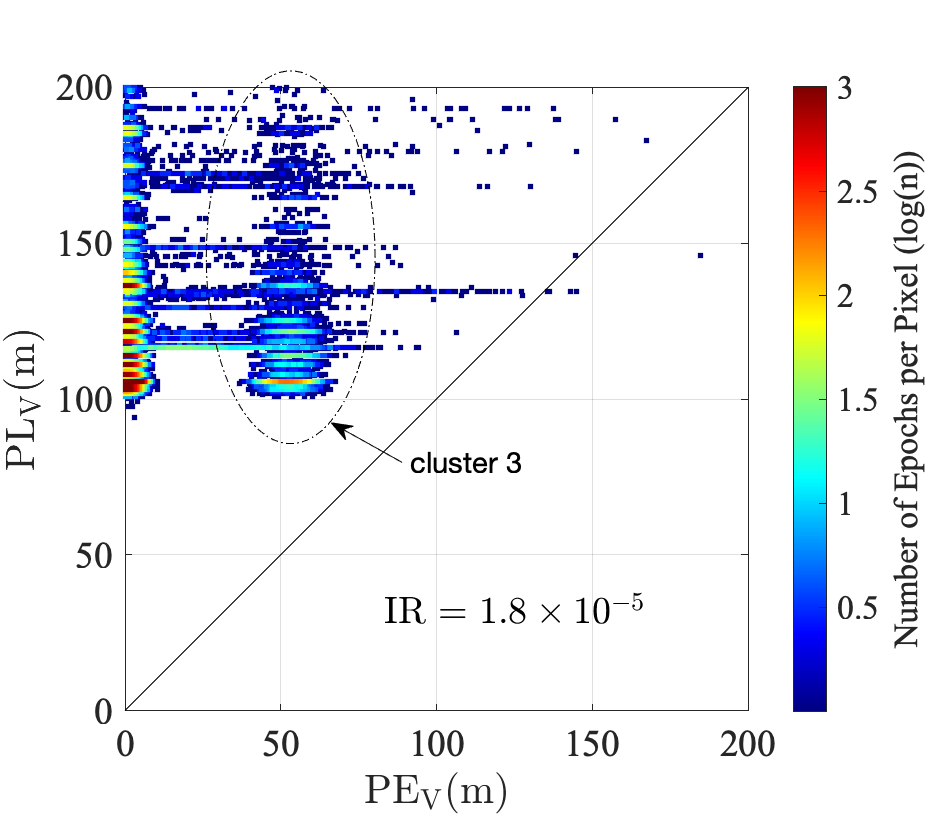}}
\caption{Performances of the RAIM algorithms in the form of Stanford diagrams under NLoS type fault conditions, with same TIR requirement $P_\mathrm{TIR} = 10^{-3}$ for all subspaces. In (a)-(d), a pixel stands for $0.01\times0.01$ m$^2$; in (e) and (f), a pixel stands for $0.2\times0.2$ m$^2$.}  \vspace{-1em}
\label{fig:SD:NloS}
\end{figure*}

\begin{figure*}[!t]
\centering\vspace{-0.5em}
 \subfigure[Bayesian RAIM, ${\mathrm{H}}$-plane Alg.~1]{ \includegraphics[width= .27\linewidth]{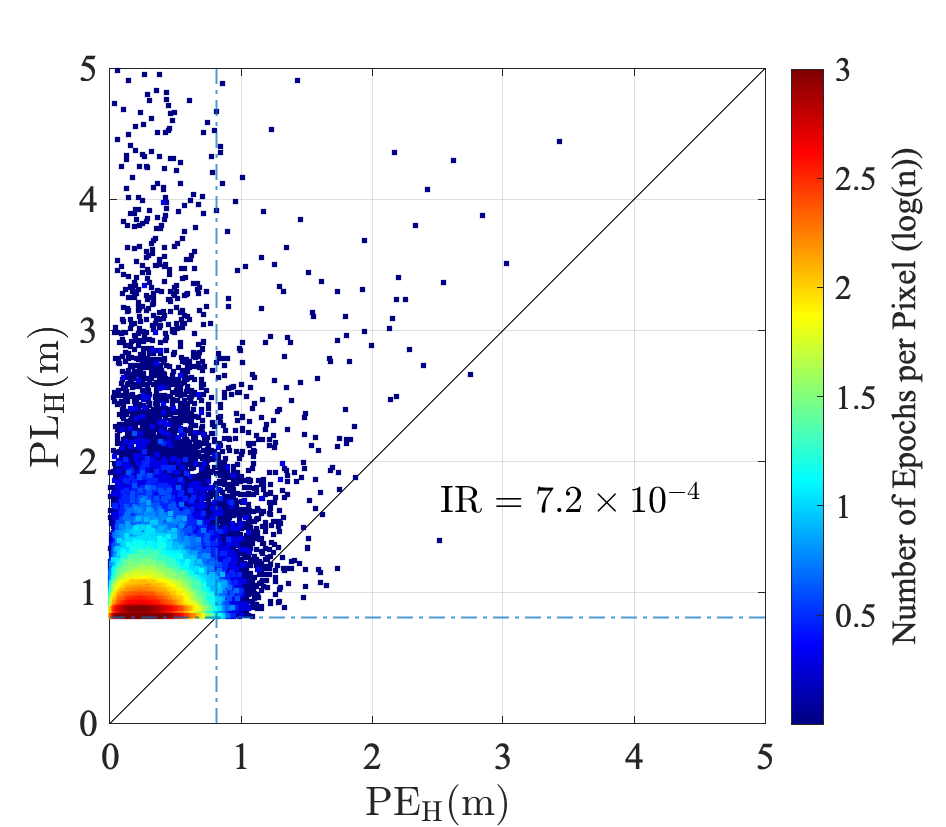}}
 \hfill
 \subfigure[Bayesian RAIM, ${\mathrm{H}}$-plane overestimate]{\includegraphics[width= .27\linewidth]{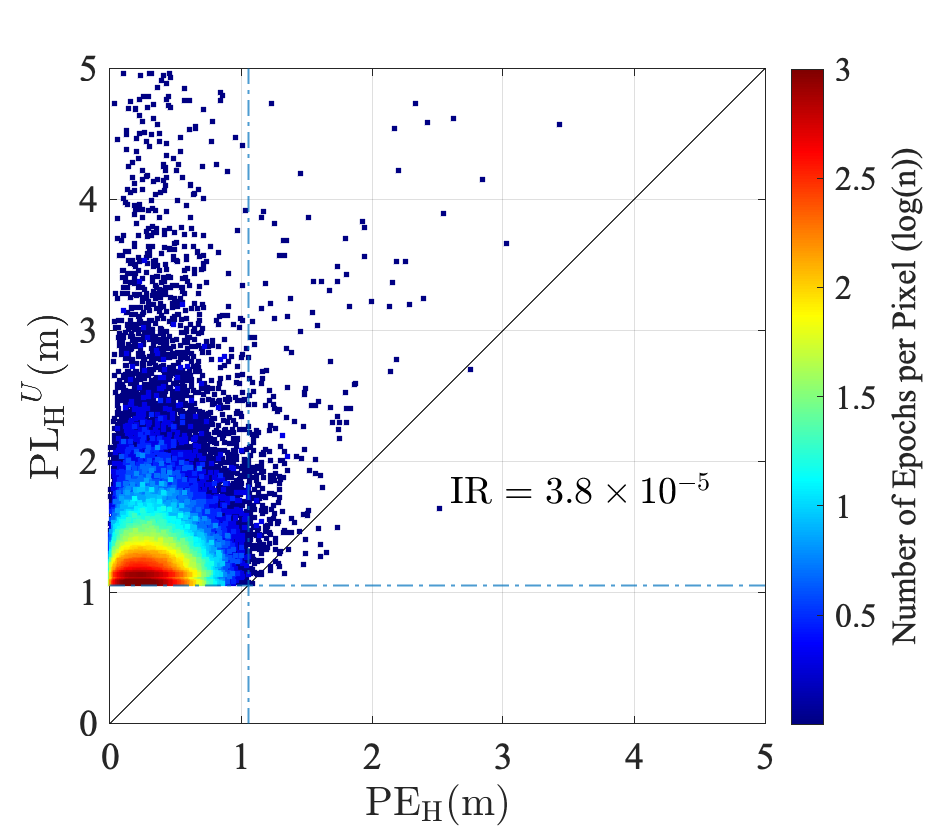}}
 \hfill\vspace{-1em}
 \subfigure[Baseline RAIM, ${\mathrm{H}}$-plane]{ \includegraphics[width= .27\linewidth]{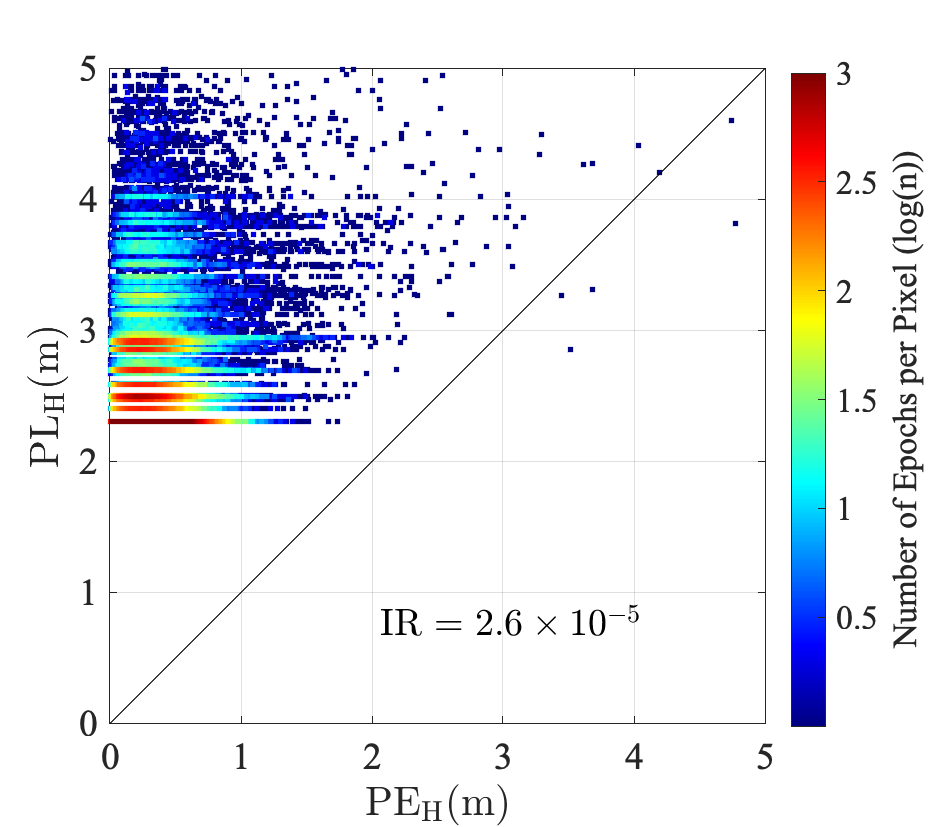}}
 \subfigure[Bayesian RAIM, $\tilde{\mathbf{v}}_{45^\circ}$-direction]{ \includegraphics[width= .27\linewidth]{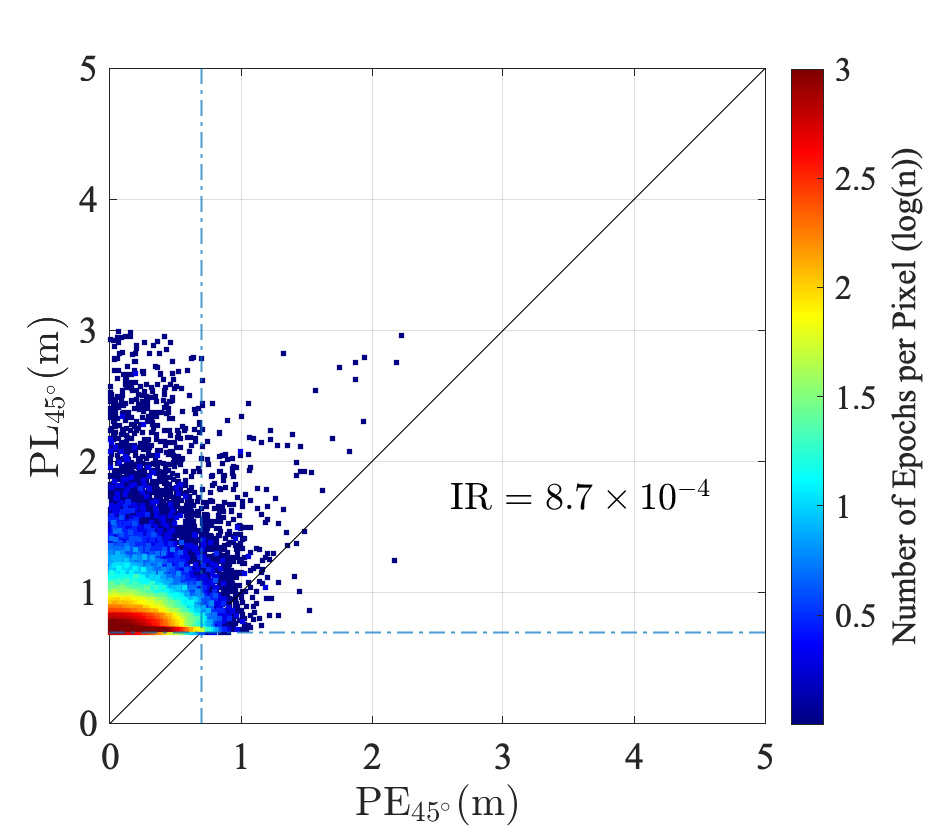}}
  \hfill
 \subfigure[Bayesian, ${\mathrm{V}}$-direction]{ \includegraphics[width= .27\linewidth]{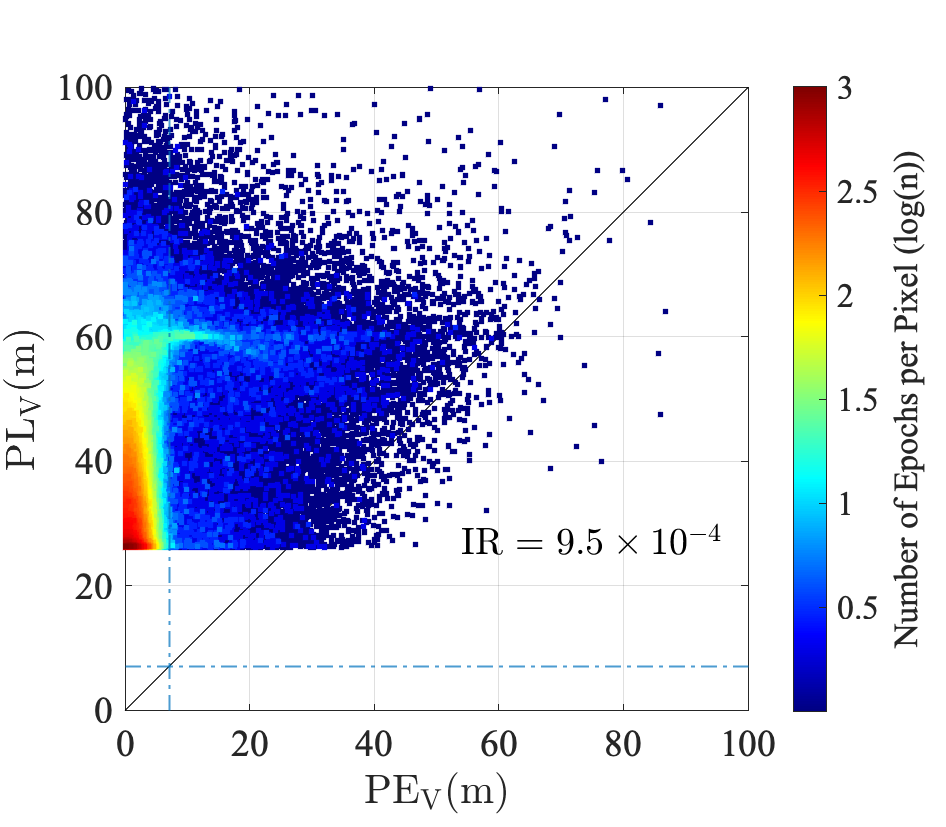}}
 \hfill\vspace{-.5em}
  \subfigure[Baseline, ${\mathrm{V}}$-direction]{ \includegraphics[width= .27\linewidth]{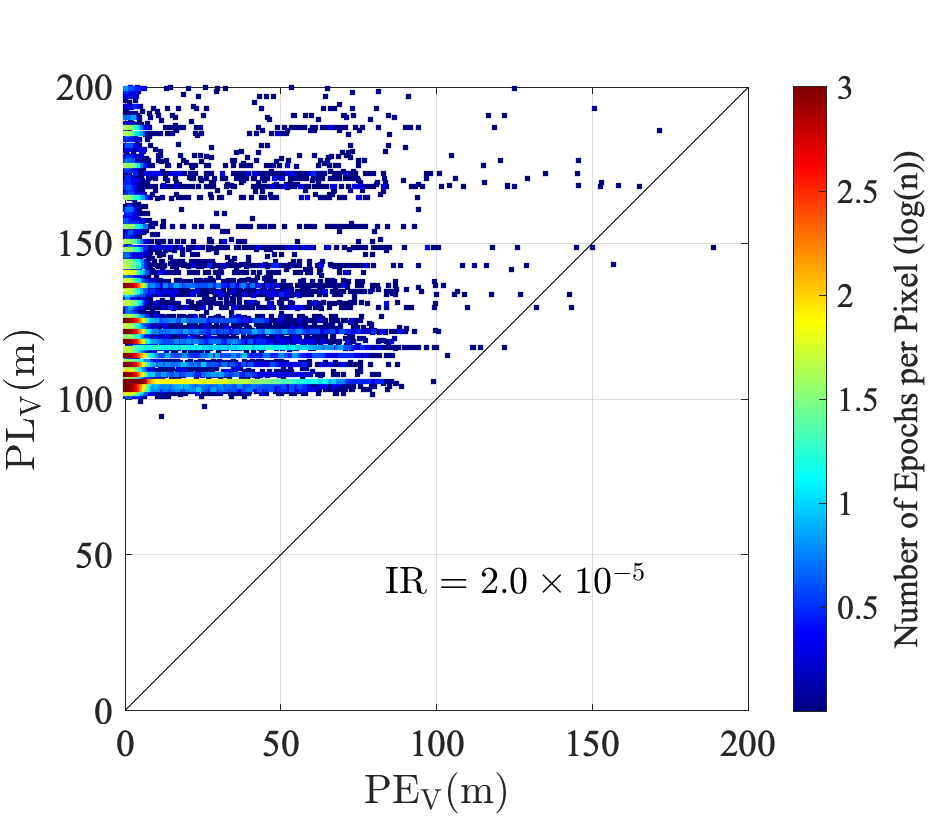}}
\caption{Performances of the RAIM algorithms in the form of Stanford diagrams under clock synchronization type fault conditions, with same TIR requirement $P_\mathrm{TIR} = 10^{-3}$ for all subspaces. In (a)-(d), a pixel stands for $0.01\times0.01$ m$^2$; in (e) and (f), a pixel stands for $0.2\times0.2$ m$^2$.}  \vspace{-1em}
\label{fig:SD:Clock}
\end{figure*}

The well-distributed BSs around the UE provide strong positioning capability on the $\mathrm{H}$-plane. In Fig.~\ref{fig:SD:NloS} and Fig.~\ref{fig:SD:Clock}, subfigures (a)-(c), Baseline RAIM achieves simulated IRs in the order of $10^{-5}$ for both fault types. Bayesian RAIM achieves similar IRs using PL overestimation and just below $P_\mathrm{TIR}$ using the exact PL computation algorithm. While fault types minimally affect Baseline RAIM, clock-type faults causes larger PLs in Bayesian RAIM, spreading the point clouds upwards due to the inseparable small bias realizations from noise. Table~\ref{tab:PL} shows that Bayesian RAIM achieved over $50\%$ PL reduction for all percentiles with the overestimation method and over $60\%$ with the exact PL computation method, except for the $99$th percentile value under clock-type faults.

\begin{table}[!t] 
\vspace{-.5em}
\caption{Empirical percentile PL values ($\mathrm{v}$) and reduction ($\mathrm{r}$, computed by $\mathrm{r}= 1 - \mathrm{PL}_\mathrm{Bayesian}/{\mathrm{PL}_\mathrm{Baseline}}$) by Bayesian RAIM compared to Baseline RAIM: \textbf{N} denotes NLoS-type fault, and \textbf{C} denotes clock-type fault.}
\label{tab:PL}
    \centering \footnotesize \vspace{-.5em}
    \begin{tabular}{|c c||c c|c c|c c|}
    \hline
        & & \multicolumn{2}{c|}{PL@$50\%$} & \multicolumn{2}{c|}{PL@$95\%$ }& \multicolumn{2}{c|}{PL@$99\%$}\\
    \cline{3-8}
        & & v[m] & r[\%] & v[m] & r[\%] & v[m] & r[\%] \\
    \hline 
        \multirow{2}{*}{\!$\mathrm{PL}_\mathrm{H}^U$\!} &\!N& $1.10$& $52.4$& $1.32$& $59.1$& $1.46$& $62.5$\\
        &\!C& $1.14$& $50.6$ & $1.44$& $51.8$& $1.80$& $51.7$ \\
    \hline
        \multirow{2}{*}{\!$\mathrm{PL}_\mathrm{H}$\!}&\!N& $0.84$& $63.5$& $1.07$& $66.7$& $1.22$& $68.7$\\
        &\!C& $0.88$& $61.7$& $1.18$& $60.5$& $1.55$& $58.4$\\
    \hline 
        \multirow{2}{*}{\!$\mathrm{PL}_\mathrm{V}$\!} &\!N& $7.32$& $93.1$& $56.53$& $58.5$& $57.48$& $65.9$\\
        &\!C& $34.12$& $67.7$& $58.63$& $56.4$& $67.81$& $55.0$\\
    \hline
    \end{tabular}\vspace{-1.em}
\end{table}

\subsubsection{\texorpdfstring{$\tilde{\mathbf{v}}_{45^\circ}$}{45}-direction} 
\begin{figure}
    \centering
    \subfigure[NLOS fault]{ \includegraphics[width= .48\linewidth]{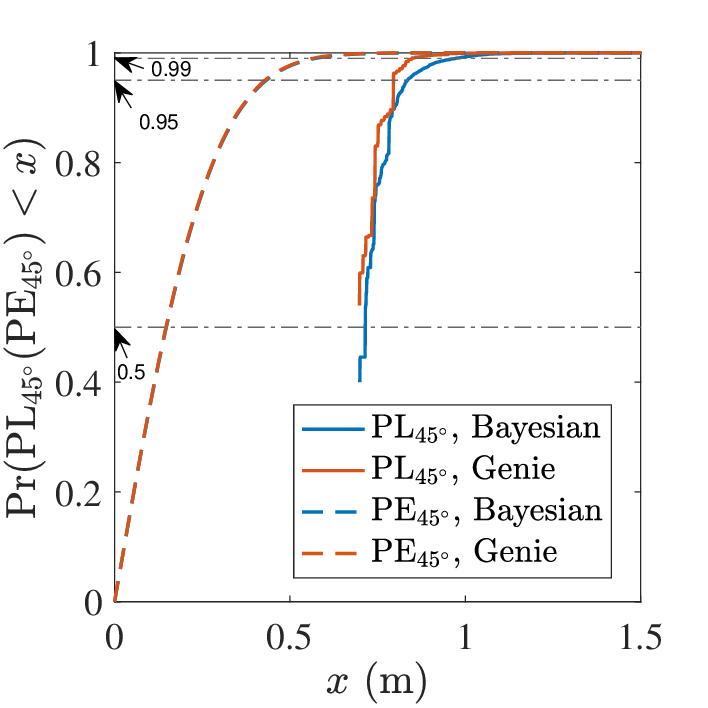}}
 \hfill
 \subfigure[Clock synchronization fault]{\includegraphics[width= .48\linewidth]{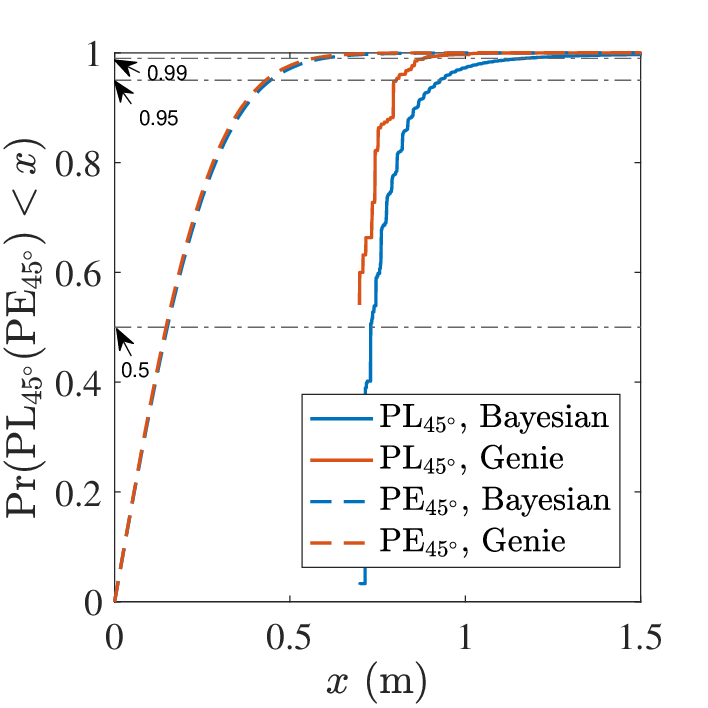}}
    \caption{Empirical CDF of PL and PE obtained by Bayesian RAIM between (1) our implementation and (2) the genie variation.}
    \label{fig:Compare:Genie}
\end{figure}

In both Fig.~\ref{fig:SD:NloS} and Fig.~\ref{fig:SD:Clock}, subfigure (d) demonstrates tighter PLs and simulated IRs closer to $P_\mathrm{TIR}$ than subfigure (a), illustrating the benefit of Bayesian RAIM when the direction of interest is known. Fig.~\ref{fig:Compare:Genie} compares the empirical CDFs of PL and PE values with those from the genie variation. The PE CDFs are nearly identical, and the PL CDFs are also very close, particularly for NLoS faults.

\subsubsection{\texorpdfstring{$\mathrm{V}$}{V}-direction}  
In both figures, subfigures (e) and (f) show Baseline RAIM achieving simulated IRs in the order of $10^{-5}$, while Bayesian RAIM remains close to $10^{-3}$. The relatively low BS heights reduce vertical positioning capability, resulting in significantly larger PL values compared to the $\mathrm{H}$-plane. Point clouds vary significantly in shape depending on the algorithm and fault type, influenced by BS geometry. Notably, they cluster into distinct groups under NLoS-type fault conditions in Fig.~\ref{fig:SD:NloS}(e) and (f), due to the proximity of the $7$th BS to the UE, making its large measurement biases influential. Specifically, in Fig.~\ref{fig:SD:NloS}(e), cluster 1 is generated by approximately $3.44\times 10^5$ epochs where the $7$th measurement is fault-free; and cluster 2 consists of $51$ epochs where the $7$th measurement is faulty but has a low posterior fault probability $\theta_i'$ \eqref{eq:theta:i:pos}, resulting in too small PLs to upper bound the PEs. In Fig.~\ref{fig:SD:NloS}(f), cluster 3 consists mainly of epochs with fault modes where the $7$th measurement is incorrectly excluded or faulty but not detected. Despite this, Bayesian RAIM consistently produces much lower PL values compared to Baseline RAIM for both fault types, with reductions ranging from  $55.0\%$ to $93.1\%$, as shown in Table~\ref{tab:PL}.

\subsubsection{Running time} 
\begin{figure}[!t]
    \centering \vspace{-.5em}
    \includegraphics[width = .99\linewidth]{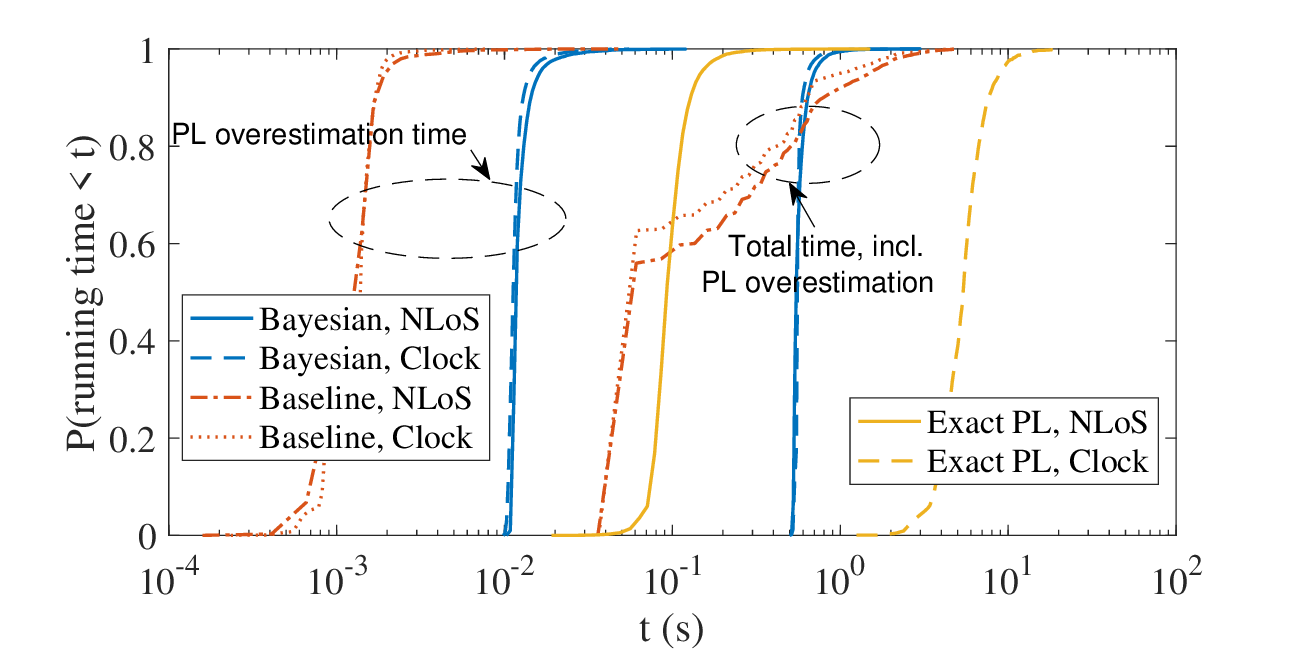} \vspace{-1.2em}
    \caption{Empirical CDF of running times, including the total and PL computation times for Baseline RAIM and Bayesian RAIM with PL overestimation, along with the running time for Algorithm 1.}
    \vspace{-0.7em}
    \label{fig:Time}
\end{figure}

Figure~\ref{fig:Time} shows empirical CDFs for the total and PL computation times of Baseline RAIM and Bayesian RAIM with PL overestimation. Fault type shows little effect on these results. Baseline RAIM’s total runtime is more variable and can exceed that of Bayesian RAIM with overestimation, which is consistently around $0.55$ seconds (median). PL overestimation adds minimal overhead to both algorithms. Figure~\ref{fig:Time} also presents CDFs for exact PL computation time, which is much longer than the overestimation method and highly dependent on fault type: about $0.1$ seconds for NLoS faults and several seconds for clock faults. This difference arises from the GM posterior PDF weights: with NLoS faults, only a few terms are significant, while clock faults involve many terms. For $\zeta_2 = 0.002$ (see \eqref{eq:constraintL}), the GM model typically reduces to fewer than 10 terms for NLoS faults, but several hundred for clock faults; even with $\zeta_2 = 0.1$, around 100 terms may remain.

\vspace{-3mm}
\subsection{Evaluation of Sensitivity to Linearization Error}
\label{sec:performance:part2}

\begin{table}[!t]
\caption{Bayesian RAIM Performance with initial positioning errors: (1) $45^\circ$-direction + NLoS-fault, (2) $45^\circ$-direction + Clock-fault, (3) V-direction + NLoS-fault, (4) V-direction + Clock-fault.}
\label{tab:Bayesian}
    \centering \footnotesize
    \begin{tabular}{|c c||c|c|c||c|c|c|}
    \hline
        & & \multicolumn{3}{c||}{PL@$50\%$ [m]} & \multicolumn{3}{c|}{Simulated IR [$\times 10^{-3}$]}\\
    \cline{3-8}
        & & min & max & CV & min & max & CV \\
    \hline 
        \multirow{4}{*}{\!$E_\mathrm{H}$\!} &1& $0.71$& $0.71$& $0.01\%$& $0.94$& $1.0$& $1.6\%$\\
        &2& $0.73$& $0.73$& $0.02\%$& $0.87$& $0.96$& $2.4\%$\\
        &3& $7.32$& $7.52$& $0.96\%$& $0.96$& $1.1$& $3.3\%$\\
        &4& $34.0$& $34.1$& $0.06\%$& $0.78$& $0.95$& $5.9\%$\\
    \hline 
        \multirow{4}{*}{\!$E_\mathrm{V}$\!} &1& $0.70$& $0.74$& $1.8\%$& $0.84$& $1.1$& $7.5\%$\\
        &2& $0.72$& $0.76$& $1.4\%$& $0.87$& $1.0$& $4.5\%$\\
        &3& $5.31$& $11.04$& $25.7\%$& $0.94$& $5.4$& $77.0\%$\\
        &4& $16.7$& $47.6$& $34.6\%$& $0.76$& $1.2$& $13.2\%$\\
    \hline
    \end{tabular}
    \vspace{-5mm}
\end{table}

\begin{table}[!t]
\caption{Baseline RAIM Performance with initial positioning errors: (1) H-plane + NLoS-fault, (2) H-plane + Clock-fault, (3) V-direction + NLoS-fault, (4) V-direction + Clock-fault.}
\label{tab:Baseline}
    \centering \footnotesize
    \begin{tabular}{|c c||c|c|c||c|c|c|}
    \hline
        & & \multicolumn{3}{c||}{PL@$50\%$ [m]} & \multicolumn{3}{c|}{Simulated IR [$\times 10^{-5}$]}\\
    \cline{3-8}
        & & min & max & CV & min & max & CV \\
    \hline 
         \multirow{4}{*}{\!$E_\mathrm{H}$\!} &1& $2.31$& $2.34$& $0.38\%$& $4.4$& $4.6$ & $2.3\%$\\
         &2& $2.31$& $2.33$& $0.35\%$& $2.4$ & $2.8 $& $5.9\%$\\
         &3& $105.5$& $106.5$& $0.32\%$& $1.6$ & $1.8$& $5.3\%$\\
         &4& $105.5$& $106.4$& $0.27\%$& $1.8$ & $2.6$& $13.7\%$\\
    \hline 
         \multirow{4}{*}{\!$E_\mathrm{V}$\!} &1& $2.20$& $2.35$& $2.4\%$& $1.8$& $5.8$ & $31.3\%$\\
         &2& $2.20$& $2.35$& $2.4\%$ & $2.6 $& $4.6 $& $17.1\%$\\
         &3& $55.5$& $148.4$& $33.6\%$& $0.4$& $3.8 $& $54.5\%$\\
         &4& $55.5$& $148.3$& $33.6\%$& $2.0 $& $3.4$& $22.5\%$\\
    \hline
    \end{tabular}
\vspace{-2mm}
\end{table}

\begin{figure}[!t]
\centering 
    \subfigure[Cases 1,2 and 4]{ \includegraphics[width= .48\linewidth]{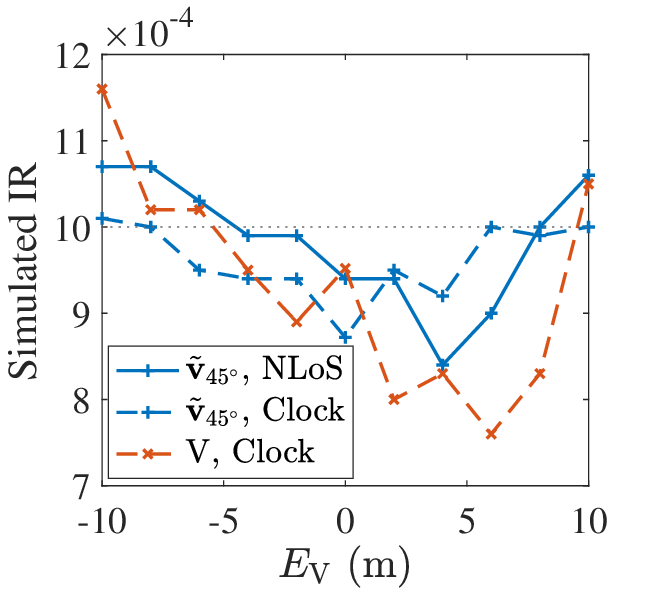}}
 \hfill
    \subfigure[Case 3]{ \includegraphics[width= .48\linewidth]{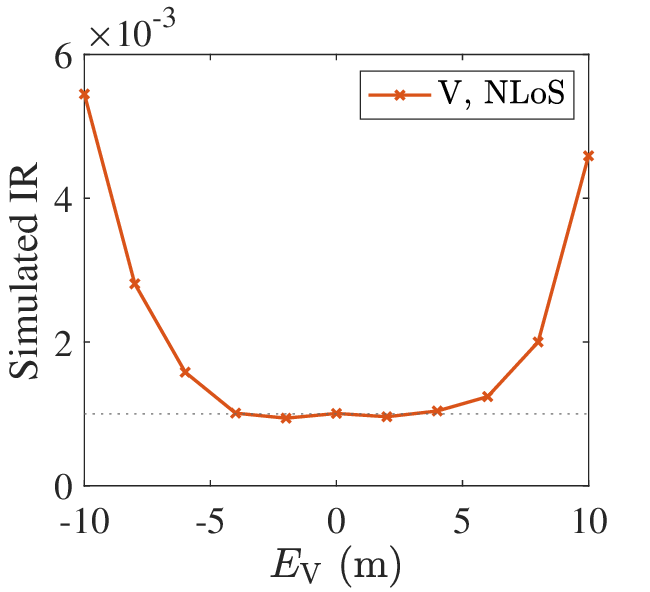}}
\caption{Simulated IR performance of Bayesian RAIM with vertical initial positioning error. 
}  \vspace{-1.5em}
\label{fig:IR_Bayesian_Verror}
\end{figure}

We evaluate the sensitivity of the algorithms to initial positioning errors using the $50$th percentile PL and the simulated IR. As the results turn show relative insensitivity as $E_\mathrm{H}$ varying in the $[0, 5]$~m range and $E_\mathrm{V}$ in $[-10,10]$~m range, we show them as tables as opposed to plots. Specifically, Table~\ref{tab:Bayesian} presents their minimum, maximum, and coefficient of variation (CV, defined as the ratio of the standard deviation to the mean) for Bayesian RAIM, while Table~\ref{tab:Baseline} provides these metrics for Baseline RAIM. 

\subsubsection{Bayesian RAIM}
As Table~\ref{tab:Bayesian} shows, the horizontal initial positioning error $E_{\mathrm{H}}$ has minimal effect on the $50$th percentile PL for both the $\mathrm{H}$-plane (cases 1-2) and the $\mathrm{V}$-direction (cases 3-4). However, it can cause the maximum simulated IR (case 3) to exceed the TIR requirement. In contrast, the vertical initial positioning error $E_{\mathrm{V}}$ significantly affects these metrics, especially in the $\mathrm{V}$-direction. Under NLoS fault conditions, the simulated IR in the $\mathrm{V}$-direction (case 3) exceeds five times the TIR. Fig.~\ref{fig:IR_Bayesian_Verror} displays the curves of the simulated IR over $E_{\mathrm{V}}$, showing that the IR increases with $|E_{\mathrm{V}}|$. No clear trends were observed for other metrics, so they are not included here.

\subsubsection{Baseline RAIM}
Table~\ref{tab:Baseline} shows that initial positioning errors similarly affect the performance of Baseline RAIM. However, because it computes much looser PLs, the simulated IRs remain in the order of $10^{-5}$. Additionally, the $50$th percentile PL values under both NLoS and Clock-type faults are almost identical, indicating the insensitivity of Baseline RAIM to the fault distribution. This contrasts with the Bayesian RAIM, which exhibits sensitivity to fault distribution.

\subsubsection{Discussion and Implications}

Bayesian RAIM computes the exact posterior \acp{PDF} of the UE’s position based on the linear measurement model \eqref{eq:linearized-3D-model} and the distributions of bias and noise, making it sensitive to model mismatches. Due to the relative low heights of the BS, the already small third entry of $\mathbf{g}_i = {(\mathbf{x}_{\mathrm u,0} - \mathbf{x}_i)}/{\Vert \mathbf{x}_{\mathrm u,0} - \mathbf{x}_i\Vert} \approx (\mathbf{x}_{\mathrm u} - \mathbf{x}_i +\mathbf{e}_{0,\mathrm{V}})/ \Vert \mathbf{x}_{\mathrm u}-\mathbf{x}_i\Vert$, is significantly affected by high vertical error magnitude $|E_\mathrm{V}|$. With an average BS height of just $19.2$ meters in our simulation, a maximum $|E_\mathrm{V}|$ value of $10$ meters can cause a substantial error in the unit vector. The results reaffirm that effective positioning performance, including robust integrity assurance, critically hinges on the geometric placement of \acp{BS}. Additionally, investigating methods to incorporate potential model errors into the problem formulation emerges as a promising direction for future research. 

\vspace{-2mm}
\section{Conclusions}
\label{sec:conclusions}

In this paper, we developed a Bayesian \ac{RAIM} algorithm for \ac{ToA}-based, snapshot 3D cellular positioning to handle measurement faults. Using message passing on a factor graph and a linearized measurement model, the algorithm efficiently computes the posterior \ac{PDF} of the \ac{UE} position. We introduced computational rules for Gaussian messages to ensure accurate scaling and robust handling of degenerate cases. This approach fully leverages all available measurements, enabling exact, low-cost computation of 1D \acp{PL} in any direction, as well as immediate overestimates for 2D or 3D \acp{PL}. We also developed an exact 2D/3D PL search by relating the probability of a Gaussian vector inside an ellipsoid to the \ac{CDF} of a generalized chi-squared variable.

For performance evaluation, we adapted an advanced RAIM algorithm \cite{blanch2015baseline} as a baseline. Monte Carlo simulations show that our Bayesian RAIM achieves over $50\%$ tighter \acp{PL} in all cases (Table~\ref{tab:PL}), even with the efficient PL overestimation approach, and closely meets the \ac{TIR} requirement with exact PL computation. PL tightness is further confirmed by comparison with the genie variation. Results also highlight the importance of good \ac{BS} geometry: under favorable conditions, Bayesian RAIM remains robust to initial position errors during linearization, making it an effective and reliable solution for 3D cellular positioning integrity.

In this paper, our primary goal is to demonstrate the Bayesian RAIM principle using simplified, idealized models as a first step. However, several extensions are possible. First, incorporating non-Gaussian noise models (e.g., Laplacian or heavy-tailed distributions) and nonlinear measurements (e.g., unlinearized \ac{ToA} and \ac{AoA}) could better capture real-world effects. Sensitivity analyses with respect to noise model mismatches would further elucidate the robustness of the proposed algorithms. Furthermore, validation using real-world datasets or synthetic datasets generated by deterministic (e.g., ray-tracing) or standardized stochastic (e.g., 3GPP TR 38.901) channel models would provide a more comprehensive assessment of performance under practical conditions. Finally, extensions to dynamic scenarios involving user and fault tracking are natural, as the factor graph framework inherently supports sequential belief updates across time steps.

\begin{appendices}
\vspace{-1.5mm}
\section{Proof of Lemma~\ref{lemma1}}
\label{App:Proof:Lemma1}

Lemma~\ref{lemma1} follows since 
\begin{align*}
& \prob{ \Vert \mathbf{e}_{n\mathrm{D}} \Vert > \mathrm{PL}_{n\mathrm{D}}^U} 
\overset{(a)}{\leq}  \prob{ \cup_{i=1}^n \, |e_i| > \mathrm{PL}_{1\mathrm{D},i} }  \nonumber \\ 
&\overset{(b)}{\leq}  \sum_{i=1}^n \prob{|e_i| > \mathrm{PL}_{1\mathrm{D},i}} 
\overset{(c)}{<} \sum_{i=1}^n w_i \, P_\mathrm{TIR}= P_\mathrm{TIR}.
\end{align*}
The inequality (a) can be verified from geometry: $\cup_{i=1}^n \, |e_i| > \mathrm{PL}_{1\mathrm{D},i}$ describes the event when the error vector $\mathbf{e}_{n\mathrm{D}}$ is located outside the 2D/3D box with edge lengths given by $\{2\mathrm{PL}_{1\mathrm{D},i}\}$; while $\Vert \mathbf{e}_{n\mathrm{D}} \Vert > \mathrm{PL}_{n\mathrm{D}}^U$ is the event when $\mathbf{e}_{n\mathrm{D}}$ is located outside the 2D circle/3D sphere with radius $\mathrm{PL}_{n\mathrm{D}}^U$, which encloses the box. The inequality (b) follows the union bound, and (c) follows from the assumption \eqref{eq:IR:split}.

\vspace{-2mm}
\section{Message Passing Rules for Problem~\ref{prob:1}}
\label{App:GMP}

\subsection{General Gaussian Distribution}

A rigorous definition of a general Gaussian random vector, following \cite[Definition 23.1.1]{lapidoth2017foundation}, is given below. An equivalent definition can be found in \cite[Definition 2.1]{nguyen2022distributionally}. 

\begin{definition}[Gaussian Random Vector]
A real random vector $\boldsymbol{X} = (X_1,\dots,X_n)^{\mathrm T}$ is said to be Gaussian if there exists a deterministic matrix $\mathsf A \in \mathbb R^{n\times m}$ and a deterministic vector $\mathbf m\in \mathbb R^n$ such that the distribution of $\boldsymbol{X}$ is equal to that of $\mathsf{A} \boldsymbol{Z} + \mathbf m$, where $\mathbf Z$ is a standard Gaussian vector with $m$ components. The distribution is denoted by $\boldsymbol{X} \sim \mathcal{N}(\mathbf{x}; \mathbf{m}, \mathsf{\Sigma})$, where $\mathsf{\Sigma} \triangleq \mathsf{AA}^{\mathrm T} \succeq 0$ is the covariance matrix. The \ac{PDF} of $\mathbf X$ is given by 
\begin{align} \label{eq:pdf_multivariate_normal} 
    f_{\boldsymbol{X}}({\mathbf{x}}) = \frac{\exp \left[-\frac{1}{2} (\mathbf{x} - \mathbf{m})^\T \mathsf{\Sigma} ^{+} (\mathbf{x} - \mathbf{m}) \right]}
    {(2\pi)^{k/2} |\mathsf{\Sigma}|_+^{1/2} }, 
\end{align}
where $k= \mathrm{rank}(\mathsf{\Sigma})$, and $\mathsf{\Sigma}^{+}$ and $|\mathsf{\Sigma}|_+$ are the pseudo-inverse\footnote{
The pseudo-inverse (Moore–Penrose inverse) of $\mathsf A$ is the matrix $\mathsf A^+$ that satisfies \cite[Section 3.6]{petersen2008matrix}: (i) $\mathsf A \mathsf A^+\mathsf A =\mathsf A$; (ii) $\mathsf A^+\mathsf A \mathsf A^+ =\mathsf A^+$; (iii) $\mathsf A \mathsf A^+$ is symmetric; (iv) $\mathsf A^+\mathsf A$ is symmetric. The pseudo-inverse is unique and always exists, with $(\mathsf A^+)^+ = \mathsf A$. 
When $\mathsf A$ has full-rank, it has explicit expressions: for full row rank, $\mathsf A^+ = \mathsf{A}^{\mathrm T}(\mathsf{A} \mathsf{A}^{\mathrm T})^{-1}$; for full column rank, $\mathsf A^+ = (\mathsf{A}^{\mathrm T} \mathsf{A})^{-1} \mathsf{A}^{\mathrm T}$.}  
and pseudo-determinant\footnote{
The pseudo-determinant of $\mathsf{\Sigma} \succeq 0$ is given by $|\mathsf{\Sigma}|_+ \triangleq \prod_{i=1}^{k} \gamma_i^+$, where $\gamma_1^+,\dots,\gamma_k^+$ are the positive eigenvalues of $\mathsf{\Sigma}$.} 
of $\mathsf{\Sigma}$, respectively. 
\end{definition}

In this definition, $n$ can be greater than, equal to, or smaller than $m$; $k\leq\min(m,n)$. When $\mathsf{\Sigma} \succ 0$, $\mathsf{\Sigma}^{+}$ and $|\mathsf{\Sigma}|_+$ coincide with the regular inverse $\mathsf{\Sigma}^{-1}$ and determinant $|\mathsf{\Sigma}|$. Expanding \eqref{eq:pdf_multivariate_normal}, we have 
\begin{align} 
    f_{\boldsymbol{X}}({\mathbf{x}}) 
    &=\frac{\exp \left(-\frac{1}{2} \mathbf{x}^\T \mathsf{\Sigma}^{+} \mathbf{x} + \mathbf{x}^\T \mathsf{\Sigma}^{+} \mathbf{m}  -\frac{1}{2}  \mathbf{m}^\T \mathsf{\Sigma}^{+} \mathbf{m}  \right)}{ {(2\pi)^{k/2}}{|\mathsf{\Sigma}|_+^{1/2}}}  \label{eq:pdf_multivariate_normal2} \\
    &= \frac{|{\mathsf{V}}|_+^{1/2} \exp \left(-\frac{1}{2} \mathbf{x}^\T \mathsf{V}  \mathbf{x} + \mathbf{x}^\T\mathbf{u} -\frac{1}{2}  \mathbf{u}^\T \mathsf{V}^{+} \mathbf{u} \right)}
    {(2\pi)^{k/2}}, 
    \label{eq:pdf_multivariate_normal3}
\end{align}
where $ \mathsf{V}  \triangleq \mathsf{\Sigma}^{+}, \quad \mathbf{u} \triangleq \mathsf{V} \mathbf{m}$, and it can be easily verified that $\mathbf{m}^\T \mathsf{\Sigma}^{+} \mathbf{m} = \mathbf{u}^\T \mathsf{V}^{+} \mathbf{u}$. This demonstrates that the Gaussian distribution can be equivalently parameterized using $(\mathbf{m}, \mathsf{\Sigma})$ or $(\mathbf{u}, \mathsf{V})$. Note that $\mathrm{rank}(\mathsf{\Sigma}) = \mathrm{rank}(\mathsf{V})$. To avoid ambiguity, we denote the \ac{PDF} expression in \eqref{eq:pdf_multivariate_normal} by $f_{\boldsymbol{X}}({\mathbf{x}}; \mathbf{m}, \mathsf{\Sigma})$ and in \eqref{eq:pdf_multivariate_normal3} by $f^{\mathrm{E}}_{\boldsymbol{X}}({\mathbf{x}}; \mathbf{u}, \mathsf{V})$. Finally, recall $\alpha_{\boldsymbol{X}} \triangleq - \frac{1}{2}  \mathbf{m}^\T \mathsf{\Sigma}^{+} \mathbf{m} \equiv  - \frac{1}{2}  \mathbf{u}^\T \mathsf{V}^{+} \mathbf{u}$, as defined in \eqref{eq:alpha}.

\vspace{-3mm}
\subsection{Proof of Lemma~\ref{lemma:InverseA}}
\label{sec:GMP:a}

Consider $\mathbf{u}_{\boldsymbol{X}}$ and $\mathsf{V}_{\boldsymbol{X}}$ given in \eqref{eq:MM:backward2}. Since $\mathsf{A}\in\mathbb{R}^{m\times n}$ has full row rank, we have $\mathrm{rank}(\mathsf{V}_{\boldsymbol{X}}) = \mathrm{rank}(\mathsf{V}_{\boldsymbol{Y}}) =m$ and $\mathsf A \mathsf A^+ =\mathsf I_m$. Therefore $ \mathbf{u}_{\boldsymbol{X}}^\T \mathsf{V}_{\boldsymbol{X}}^{+} \mathbf{u}_{\boldsymbol{X}} 
    = \mathbf{u}_{\boldsymbol{Y}}^\T {\mathsf{A}} \mathsf{A}^+ \mathsf{V}_{\boldsymbol{Y}}^+  (\mathsf{A}\mathsf{A}^+)^\T \mathbf{u}_{\boldsymbol{Y}} = \mathbf{u}_{\boldsymbol{Y}}^\T \mathsf{V}_{\boldsymbol{Y}}^+\mathbf{u}_{\boldsymbol{Y}}$, where the property $(\mathsf{AB})^+ = \mathsf{B}^+\mathsf{A}^+$ is used. Thus, $\alpha_{\mathbf X}= \alpha_{\mathbf Y}$. Given this, we confirms the correctness of \eqref{eq:MM:backward1}--\eqref{eq:MM:backward3}:
\begin{align*}
    &f^{\mathrm{E}}_{\boldsymbol{Y}}({\mathsf{A}\mathbf{x}}; \mathbf{u}_{\boldsymbol{Y}}, \mathsf{V}_{\boldsymbol{Y}}) \\
    &= \frac{|\mathsf{V}_{\boldsymbol{Y}}|_+^{1/2}}{(2\pi)^{m/2}}
    {\exp \left[-\frac{1}{2} (\mathsf{A}\mathbf{x})^\T \mathsf{V}_{\boldsymbol{Y}}  \mathsf{A}\mathbf{x} + (\mathsf{A}\mathbf{x})^\T \mathbf{u}_{\boldsymbol{Y}} +\alpha_{\mathbf Y} \right]} \\
    &= \underbrace{\frac{ |{\mathsf{V}}_{\boldsymbol{Y}}|_+^{1/2} }{|\mathsf{V}_{\boldsymbol{X}}|_+^{1/2}} }_{s^{\boldsymbol{X}}_{\boldsymbol{Y}} }
    \underbrace{ \frac{ |\mathsf{V}_{\boldsymbol{X}}|_+^{1/2}}{(2\pi)^{m/2}}
    {\exp \Big(-\frac{1}{2} \mathbf{x}^\T \mathsf{V}_{\boldsymbol{X}} \mathbf{x} +
    \mathbf{x}^\T \mathbf{u}_{\boldsymbol{X}} + \alpha_{\mathbf X} \Big)} }_{f_{\boldsymbol{X}}^{\mathrm{E}}({\mathbf{x}}; \mathbf{u}_{\boldsymbol{X}}, \mathsf{V}_{\boldsymbol{X}})}. 
\end{align*}  

\vspace{-3mm}
\subsection{Proof of Lemma~\ref{lemma:Gaussian:product}}
\label{sec:GMP:b}

The following lemma is need for the proof of Lemma~\ref{lemma:Gaussian:product}. It provides the computation rule for the product of two  arbitrary, possibly degenerate, Gaussian densities. 

\begin{lemma}[Product of two Gaussian densities]
\label{lemma:product:2}
    Given two messages of $\boldsymbol{X}$ in the form of Gaussian densities: ${\mu}_{\boldsymbol{X},1}(\mathbf{x}) = f^{\mathrm{E}}_{\boldsymbol{X}}({\mathbf{x}}; \mathbf{u}_{\boldsymbol{X}_1}, \mathsf{V}_{\boldsymbol{X}_1})$ and ${\mu}_{\boldsymbol{X},2}(\mathbf{x}) = f^{\mathrm{E}}_{\boldsymbol{X}}({\mathbf{x}}; \mathbf{u}_{\boldsymbol{X}_2}, \mathsf{V}_{\boldsymbol{X}_2})$, where $\mathrm{rank}( \mathsf{V}_{\boldsymbol{X}_1}) =k_1 $ and $\mathrm{rank}(\mathsf{V}_{\boldsymbol{X}_2}) =k_2$, their product is given by
\begin{subequations}
\begin{align}
    f^{\mathrm{E}}_{\boldsymbol{X}}({\mathbf{x}}; \mathbf{u}_{\boldsymbol{X}_1}, \mathsf{V}_{\boldsymbol{X}_1})  f^{\mathrm{E}}_{\boldsymbol{X}}({\mathbf{x}}; \mathbf{u}_{\boldsymbol{X}_2}, \mathsf{V}_{\boldsymbol{X}_2}) 
    = s_{\boldsymbol{X}_{1:2}} f^{\mathrm{E}}_{\boldsymbol{X}}({\mathbf{x}}; \mathbf{u}_{\boldsymbol{X}}, \mathsf{V}_{\boldsymbol{X}}), \label{eq:3way:1}
\end{align}
where $\mathbf{u}_{\boldsymbol{X}} = \mathbf{u}_{\boldsymbol{X}_1} + \mathbf{u}_{\boldsymbol{X}_2}$ and $  \mathsf{V}_{\boldsymbol{X}} = \mathsf{V}_{\boldsymbol{X}_1} + \mathsf{V}_{\boldsymbol{X}_2}$, and the scaling factor $s_{\boldsymbol{X}_{1:2}}$ is given by ($k=\mathrm{rank}( \mathsf{V}_{\boldsymbol{X}})$)
\begin{align}
\label{eq:3way:3}
    s_{\boldsymbol{X}_{1:2}}  =  \frac{|\mathsf{V}_{\boldsymbol{X}_1}|_+^{1/2} {|\mathsf{V}_{\boldsymbol{X}_2}|_+^{1/2}} }{{(2\pi)^{(k_1+k_2-k)/2}} {|\mathsf{V}_{\boldsymbol{X}}|_+^{1/2}}}   
  \exp(\alpha_{\boldsymbol{X}_1} + \alpha_{\boldsymbol{X}_2} - \alpha_{\boldsymbol{X}} ).
\end{align}
\end{subequations} 
\end{lemma}

\begin{proof}
Using the \ac{PDF} expression \eqref{eq:pdf_multivariate_normal3}, we immediately have 
\begin{align*}
   & f^{\mathrm{E}}_{\boldsymbol{X}}({\mathbf{x}}; \mathbf{u}_{\boldsymbol{X}_1}, \mathsf{V}_{\boldsymbol{X}_1})  f^{\mathrm{E}}_{\boldsymbol{X}}({\mathbf{x}}; \mathbf{u}_{\boldsymbol{X}_2}, \mathsf{V}_{\boldsymbol{X}_2})  
    =\frac{{|\mathsf{V}_{\boldsymbol{X}_1}|_+^{1/2}} }{{(2\pi)^{k_1/2}} }
    \frac{ {|\mathsf{V}_{\boldsymbol{X}_2}|_+^{1/2}} }{{(2\pi)^{k_2/2}} } \cdot \\
    &{\exp \left[-\frac{1}{2} \mathbf{x}^\T (\mathsf{V}_{\boldsymbol{X}_1} \!+\! \mathsf{V}_{\boldsymbol{X}_2}) \mathbf{x} 
    + \mathbf{x}^\T (\mathbf{u}_{\boldsymbol{X}_1}\!+\! \mathbf{u}_{\boldsymbol{X}_2} ) +\alpha_{\boldsymbol{X}_1} \!+\! \alpha_{\boldsymbol{Y}_2}  \right]} \\
    =\, & \underbrace{\frac{{|\mathsf{V}_{\boldsymbol{X}_1}|_+^{1/2}} {|\mathsf{V}_{\boldsymbol{X}_2}|_+^{1/2}} }{{(2\pi)^{(k_1+k_1-k)/2}} {|\mathsf{V}_{\boldsymbol{X}}|_+^{1/2}}}
    \exp(\alpha_{\boldsymbol{X}_1} + \alpha_{\boldsymbol{X}_2} - \alpha_{\boldsymbol{X}})}_{s_{\boldsymbol{X}_{1:2}}} \cdot \\
    &\underbrace{\frac{{|\mathsf{V}_{\boldsymbol{X}}|_+^{1/2}}}{{(2\pi)^{k/2}}}
    \exp \left(-\frac{1}{2} \mathbf{x}^\T \mathsf{V}_{\boldsymbol{X}}  \mathbf{x} + \mathbf{x}^\T \mathbf{u}_{\boldsymbol{X}} +\alpha_{\boldsymbol{X}}  \right)}_{f^{\mathrm{E}}_{\boldsymbol{X}}({\mathbf{x}}; \mathbf{u}_{\boldsymbol{X}}, \mathsf{V}_{\boldsymbol{X}})}.
\end{align*}  
If $\mathsf{V}_{\boldsymbol{X}_1} \succ 0$ and $\mathsf{V}_{\boldsymbol{X}_2} \succ 0$, \eqref{eq:3way:1}--\eqref{eq:3way:3} can be obtained by using \cite[Eq.~(358)-Eq.~(364)]{petersen2008matrix}. 
\end{proof}
Based Lemma~\ref{lemma:product:2}, the computation rule for the product of multiple Gaussian densities in Lemma~\ref{lemma:Gaussian:product} can be verified by successively performing the product of two Gaussian densities. 

\vspace{-2mm}
\section{Message Passing Algorithm}
\label{App:Message:Passing}

In the following description, a Gaussian random variable is treated as a special case ($n=1$) of a Gaussian random vector and the same notations are used. 

\textbf{Step} \circled{1}: The message $p_{\Lambda_i}(\lambda_i)$ is sent from the leaf node to the variable node $\Lambda_i$ and then directly to $F_i$.

\textbf{Step} \circled{2}: Factor node $F_i$ sends the following message, 1D \ac{GM} consisting of two terms, to variable node $B_i$: 
    \begin{align}
        \mu&_{F_i\rightarrow B_i}(b_i) = \sum_{ \lambda_i=0,1} p_{B_i \mid \Lambda_i=\lambda_i}(b_i)\,p_{\Lambda_i}(\lambda_i) \nonumber\\
        &= (1-\theta_i) \, f_{B_i}(b_i;0,0) + \theta_i \, f_{B_i}(b_i;m_{b,i},\sigma_{b,i}^2).
    \end{align}
   This prior \ac{PDF} of $B_i$ is passed directly to factor node $G_i$. 
   
\textbf{Step} \circled{3}: Factor node $G_i$ sends the following message to variable node $\Gamma_i$: 
    \begin{align*}
        \mu_{G_i \rightarrow \Gamma_i} (\gamma_i) 
        = \int G_i\cdot \mu_{F_i \rightarrow B_i} (b_i) \, \mathrm{d} b_i ,
    \end{align*} 
    where $G_i = p_{Y_i \mid \Gamma_i=\gamma_i, B_i=b_i}(y_i) = f_{Y_i}(y_i; \gamma_i+b_i, \sigma_{n,i}^2) = f_{B_i}(b_i; y_i-\gamma_i, \sigma_{n,i}^2)$. Following \cite[Eqs.~(15)-(16)]{ding2022bayesian}, the integration results in a 1D \ac{GM} density with two terms for $\Gamma_i$:
    \begin{align}
        \mu_{G_i \rightarrow \Gamma_i} (\gamma_i) =\, 
        &(1-\theta_i) f_{\Gamma_i}(\gamma_i; y_i,\sigma_{n,i}^2) \nonumber\\
        &+\theta_i f_{\Gamma_i}(\gamma_i; y_i\!-m_{b,i},\sigma_{n,i}^2\!+\!\sigma_{b,i}^2).
    \end{align} 
    This message is passed directly to factor node $K_i$.

\textbf{Step} \circled{4}: Factor node $K_i$ sends the following message to variable node $\boldsymbol{X}$: 
    \begin{align*}
        \mu_{K_i \rightarrow \boldsymbol{X}} (\mathbf{x}) &= \int \delta(\gamma_i - \mathbf{h}_i^{\mathrm T} \mathbf{x})\cdot\mu_{G_i \rightarrow \Gamma_i} (\gamma_i) \, \mathrm{d} \gamma_i.
    \end{align*}
    Applying the computations rules \eqref{eq:MM:backward1}, \eqref{eq:MM:backward2}, and \eqref{eq:MM:backward3:replacement}, 
    \begin{align}\label{eq:Ki:to:X}
        \mu_{K_i \rightarrow \boldsymbol{X}}(\mathbf{x}) \propto
        \sum_{l=1,2} w^{(l)}_{\boldsymbol{X}_i} 
        \, f_{\boldsymbol{X}}^{\mathrm E}(\mathbf{x}; {\mathbf{u}}^{(l)}_{\boldsymbol{X},i},{\mathsf{V}}\!\,^{(l)}_{\boldsymbol{X}_i})
    \end{align} 
    where 
    \begin{subequations}
    \begin{align}
        &{w}_{\boldsymbol{X}_i}^{(1)} = \frac{1-\theta_i}{\sigma_{n,i} }, \quad {w}_{\boldsymbol{X}_i}^{(2)} = \frac{\theta_i}{({\sigma_{n,i}^2+\sigma_{b,i}^2})^{1/2}},  \\
        &{\mathbf{u}}_{\boldsymbol{X}_i}^{(1)} = \frac{y_i}{\sigma_{n,i}^2} \mathbf{h}_i, \quad 
        {\mathbf{u}}_{\boldsymbol{X}_i}^{(2)} = \frac{y_i-m_{b,i}}{\sigma_{n,i}^2+\sigma_{b,i}^2} \mathbf{h}_i \\
        &{\mathsf{V}}\!\,^{(1)}_{\boldsymbol{X}_i} = \frac{1}{\sigma_{n,i}^2}  \mathbf{h}_i \mathbf{h}_i^{\T}, \quad 
        {\mathsf{V}}\!\,^{(2)}_{\boldsymbol{X}_i} = \frac{1}{\sigma_{n,i}^2+\sigma_{b,i}^2}  \mathbf{h}_i \mathbf{h}_i^{\T}, 
    \end{align}
    \end{subequations} 
    Since the $\alpha$ parameter of $\boldsymbol{X}_i$ is the same as that of $K_i$, we have   
    \begin{align}\label{eq:alpha_Xip}
        {\alpha}_{\boldsymbol{X}_i}^{(1)} = -\frac{1}{2} \frac{y_i^2}{\sigma_{n,i}^2},\quad {\alpha}_{\boldsymbol{X}_i}^{(2)} = -\frac{1}{2} \frac{(y_i-m_{b,i})^2}{\sigma_{n,i}^2+\sigma_{b,i}^2}
    \end{align}
    The GM model \eqref{eq:Ki:to:X} for $\boldsymbol{X}\in \mathbb{R}^4$ consists of two degenerate Gaussian densities in a 1D subspace.

\textbf{Step} \circled{5}: Node $\boldsymbol{X}$ computes the product of all the messages sent to it: $\mu_{\boldsymbol{X}} (\mathbf{x}) \propto \prod_{j=1}^M  \mu_{K_j \rightarrow \boldsymbol{X}} (\mathbf x)$. Since each message is a GM of two terms, the result is a GM of $L \triangleq  2^{M}$ terms: 
    \begin{align}
        \mu_{\boldsymbol{X}} (\mathbf{x}) 
        & \propto \prod\nolimits_{j=1}^M \left( \sum\nolimits_{l_j=1,2} w^{(l_j)}_{\boldsymbol{X}_j} \, f_{\boldsymbol{X}}^{\mathrm E}(\mathbf{x}; {\mathbf{u}}^{(l_j)}_{\boldsymbol{X}_j},{\mathsf{V}}\!\,^{(l_j)}_{\boldsymbol{X}_j}) \right) \nonumber \\
        &\propto  \sum\nolimits_{l=1}^{L} w^{(l)}_{\boldsymbol{X}} 
        f_{\boldsymbol{X}}^{\mathrm E}(\mathbf{x}; {\mathbf{u}}^{(l)}_{\boldsymbol{X}},{\mathsf{V}}\!\,^{(l)}_{\boldsymbol{X}}) .
        \label{eq:MP:step5}
    \end{align}
    Each term of \eqref{eq:MP:step5} is computed as following: For $l_j \in \{ 1,2\}$, $j=1,\ldots, M$, $l \gets 1+\sum_{j=1}^M (l_j-1)\,2^{j-1}$, and 
    \begin{align*}
        w^{(l)}_{\boldsymbol{X}} 
        f_{\boldsymbol{X}}^{\mathrm E}(\mathbf{x}; {\mathbf{u}}^{(l)}_{\boldsymbol{X}}{\mathsf{V}}\!\,^{(l)}_{\boldsymbol{X}}) \propto \prod\nolimits_{j=1}^M w^{(l_j)}_{\boldsymbol{X}_j} 
        \, f_{\boldsymbol{X}}^{\mathrm E}(\mathbf{x}; {\mathbf{u}}^{(l_j)}_{\boldsymbol{X}_j},{\mathsf{V}}\!\,^{(l_j)}_{\boldsymbol{X}_j}),
    \end{align*}
    where, following \eqref{eq:PMG:1}, \eqref{eq:PMG:2}, and \eqref{eq:PMG:3:replacement}, 
    \begin{subequations}
    \begin{align}\label{eq:MP:step5:a}
        \mathbf{u}_{\boldsymbol{X}}^{(l)} = \sum\nolimits_{j=1}^M\mathbf{u}_{\boldsymbol{X}_j}^{(l_j)} , \quad  
        \mathsf{V}_{\boldsymbol{X}}^{(l)} = \sum\nolimits_{j=1}^M\mathsf{V}_{\boldsymbol{X}_j}^{(l_j)}, 
    \end{align}
    and 
    \begin{align}\label{eq:MP:step5:b}
        w^{(l)}_{\boldsymbol{X}}  = s^{(l)}_{\boldsymbol{X}} \prod\nolimits_{j=1}^M w^{(l_j)}_{\boldsymbol{X}_j}  ,
    \end{align}
    where, after omitting the constant $(2\pi)^{(\sum_{i=1}^K k_{i} -k)/2}$ in \eqref{eq:PMG:3:replacement}, the scaling factor is given by
    \begin{align}\label{eq:MP:step5:c}
    s^{(l)}_{\boldsymbol{X}} =  \frac{1}{{|\mathsf{V}^{(l)}_{\boldsymbol{X}}|_+^{1/2}}}   
  \exp\Big(- \alpha^{(l)}_{\boldsymbol{X}}+\sum\nolimits_{j=0}^M \alpha_{\boldsymbol{X}_j}^{(l_j)} \Big).
    \end{align}
    \end{subequations}

\textbf{Step} \circled{6}: Variable node $\boldsymbol{X}$ sends a message $\mu_{\boldsymbol{X} \rightarrow K_i} (\mathbf{x}) \propto \prod_{j\neq i}  \mu_{K_j \rightarrow \boldsymbol{X}} (\mathbf x)$ to factor node $K_i$. The computation follows the same rules as step \circled{5} and result in a GM of $2^{M-1}$ terms.

\textbf{Step} \circled{7}: Factor node $K_i$ converts $\mu_{\boldsymbol{X} \rightarrow K_i} (\mathbf{x})$ to $\mu_{K_i \rightarrow \Gamma_i} (\gamma_i)$, a 1D GM model of $\Gamma_i$. 
Each term of $\mu_{\boldsymbol{X} \rightarrow K_i} (\mathbf{x})$ is converted to a Gaussian density of $\Gamma_i$ following the linear mapping\footnote{The Gaussian message passing rule for the linear mapping $\boldsymbol{Y} = \mathsf{A}\boldsymbol{X}$ is trivial. Given ${\mu}_{\boldsymbol{X}}(\mathbf{x}) = f_{\boldsymbol{X}}({\mathbf{x}}; \mathbf{m}_{\boldsymbol{X}}, \mathsf{\Sigma}_{\boldsymbol{X}})$, the inferred message for $\boldsymbol{Y}$ is ${\mu}_{\boldsymbol{Y}}(\mathbf{y}) = f_{\boldsymbol{Y}}({\mathbf{y}}; \mathbf{m}_{\boldsymbol{Y}}, \mathsf{\Sigma}_{\boldsymbol{Y}})$ where $\mathbf{m}_{\boldsymbol{Y}} = \mathsf{A} \mathbf{m}_{\boldsymbol{X}}$ and $\mathsf{\Sigma}_{\boldsymbol{Y}} = \mathsf{A} \mathsf{\Sigma}_{\boldsymbol{X}} \mathsf{A}^{\T}$.} $\Gamma_i = \mathbf{h}_i^{\mathrm T} \boldsymbol{X}$, while retaining its weight. This message is sent to $\Gamma_i$ and then directly to factor node $G_i$.

\textbf{Step} \circled{8}: Factor node $G_i$ sends $\mu_{G_i \rightarrow B_i} (b_i)$, a 1D GM model of $2^{M-1}$ terms, to variable node $B_i$ and then to $F_i$. 

\textbf{Step} \circled{9}: Factor node $F_i$ sends a message $\mu_{F_i \rightarrow \Lambda_i} (\lambda_i)$, which is the posterior \ac{PMF} of $\Lambda_i$ after normalization, to variable node $\Lambda_i$. Specifically, the posterior probability of measurement $y_i$ being faulty is given by 
    \begin{align}\label{eq:theta:i:pos}
        \theta_i' \triangleq 
        \frac{\mu_{F_i \rightarrow \Lambda_i} (\lambda_i=1)}{\mu_{F_i \rightarrow \Lambda_i} (\lambda_i=0)+\mu_{F_i \rightarrow \Lambda_i} (\lambda_i=1)}.
    \end{align}

\vspace{-2mm}
\section{Proof of Theorem~\ref{theorem1}}
\label{App:Ellip:PR}

We first show that the probability of a Gaussian distributed vector in $\mathbb R^N$ lying within an arbitrary ellipsoid in $\mathbb R^N$ can be calculated by computing the \ac{CDF} of a generalized chi-squared random variable. An ellipsoid in $\mathbb{R}^N$ centered at $\mathbf{c} \in \mathbb{R}^N$ is defined as $ \mathcal{E}(\mathsf{A}, \mathbf{c}, \rho)=  \left\{ \mathbf{x}\in\mathbb{R}^N: (\mathbf{x} - \mathbf{c} )^\T \mathsf{A}(\mathbf{x} - \mathbf{c}) \leq \rho \right\}$, where $\mathsf{A} \in \mathbb{R}^{N\times N}$ is symmetric positive definite, and $\rho$ is a positive real value. The eigendecomposition of $\mathsf{A}$ is $\mathsf{A}= \mathsf{Q} \mathsf{D} \mathsf{Q}^\T$, where $\mathsf{D} = \mathrm{diag}(d_1,\dots,d_N)$ and $\mathsf{Q} \in \mathbb{R}^{N\times N}$ is orthogonal. The lengths of the $N$ semi-axes are given by $\rho/\sqrt{d_1}, ..., \rho/\sqrt{d_N}$. 
Consider a random vector $\boldsymbol{X} \in \mathbb{R}^N$ that follows a Gaussian distribution $\mathcal{N}(\mathbf{x};\mathbf{m},\mathsf{\Sigma})$ where $\mathsf{\Sigma}\succ 0$. Let $Z = (\boldsymbol{X} - \mathbf{c} )^\T \mathsf{A}(\boldsymbol{X} - \mathbf{c})$. The probability of $\boldsymbol{X}$ lying within $\mathcal{E}(\mathsf{A}, \mathbf{c}, \rho)$ is 
\begin{align}
    \int_{\mathcal{E}(\mathsf{A}, \mathbf{c},\rho)}  f_{\boldsymbol{X}}(\mathbf{x}; \mathbf{m},\mathsf{\Sigma}) \mathrm{d}\mathbf{x} = F_Z(\rho),
\end{align}
where $F_Z(z) \triangleq \mathrm{Pr}(Z\leq z)$ is the \ac{CDF} of $Z$. We will show that $Z$ is a generalized chi-squared random variable determined by $\mathbf{m}$, $\mathsf{\Sigma}$, $\mathsf{A}$, $\mathbf{c}$. 

Given $\mathsf{A}\succ 0$ and $\mathsf{\Sigma} \succ 0$, we have  
\begin{align}\label{eq:EigenDecomp_Y}
     \mathsf{\Sigma}^{\frac{1}{2}} \mathsf{A} \mathsf{\Sigma}^{\frac{1}{2}} = \mathsf{P} \mathsf{\Omega} \mathsf{P}^\T,
\end{align}
where $\mathsf{\Omega} = \mathrm{diag}(\omega_1,\dots,\omega_N)$ with $\omega_i >0$, and $\mathsf{P} \in \mathbb{R}^{N\times N}$ is orthogonal. It can be shown that $\boldsymbol{Y} = \mathsf{P}^\T \mathsf{\Sigma}^{-\frac{1}{2}} (\boldsymbol{X}-\mathbf{c}) \sim \mathcal{N}(\mathbf{y}; \boldsymbol{\nu} ,\mathsf{I}_N)$, where 
\begin{align}\label{eq:nu}
    \boldsymbol{\nu} \triangleq \mathsf{P}^\T \mathsf{\Sigma}^{-\frac{1}{2}} (\mathbf{m}-\mathbf{c}).
\end{align}
Letting $Y_i$ be the $i$-th element of $\boldsymbol{Y}$ and $\nu_i$ be the $i$-th entry of $\boldsymbol{\nu}$. Then $Y_i^2$ follows a noncentral chi-square distribution with one degree of freedom and noncentrality parameter $\nu_i^2$ (i.e., $Y_i^2 \sim \chi^2\big(1, \nu_{i}^2\big)$). We can rewrite $Z$ as 
\begin{align}\label{eq:weightedY}
    Z = \boldsymbol{Y}^\T \mathsf{\Omega} \boldsymbol{Y} 
    = \sum_{i=1}^{N} \omega_i Y_i^2.
\end{align}
This shows that $Z$ is a weighted sum of $N$ independent noncentral chi-square random variables. Summarizing the above, we have the following lemma. 
\begin{lemma}\label{lemma2}
    The probability that a Gaussian random vector  $\boldsymbol{X} \in \mathbb{R}^N$ $\sim\mathcal{N}(\mathbf{x};\mathbf{m},\mathsf{\Sigma})$ lies within an ellipsoid $\mathcal{E}(\mathsf{A}, \mathbf{c},\rho) \in \mathbb{R}^N$ is given by $F_Z(\rho)$, the CDF of a generalized chi-squared random variable $Z$ evaluated at $\rho$. Specifically, $Z = \sum_{i=1}^{N} \omega_i W_i$ where $W_i \sim \chi^2\big(1, \nu_{i}^2\big)$, $i=1\ldots, N$, and the weights $\{\omega_{i}\}$ and noncentrality parameters $\{\nu_{i}^2\}$ are determined by $\mathbf{m}$, $\mathsf{\Sigma}$, $\mathsf{A}$, $\mathbf{c}$ via \eqref{eq:EigenDecomp_Y} and \eqref{eq:nu}. 
\end{lemma}

By setting $\mathsf{A} = \mathsf{I}_n$, $\mathbf{c}=\mathbf{0}$, and $\rho=r^2$, the ellipsoid $\mathcal{E}(\mathsf{A}, \mathbf{c},\rho)$ becomes a circle (for $n=2$) or an sphere (for  $n=3$) centered at the origin with radius $r$. Using the \acp{PDF} of $\mathbf{e}_{n\mathrm{D}}$ given by \eqref{eq:pdf:Error:3D} and \eqref{eq:pdf:Error:1D2D}, the probability that $\mathbf{e}_{n\mathrm{D}}$ lies within a circle/sphere of radius $r$ is $\prob{\mathbf e_{n\mathrm{D}} \in \mathcal{E}(\mathsf{I}_n, \mathbf{0}, r^2) } = \sum_{l=1}^{L}  w^{(l)}_{\boldsymbol{X}} F_{Z_l}(r^2)$, where $F_{Z_l}(r^2)$ is the CDF of the generalized chi-squared random variable $Z_l$ given by  \eqref{eq:Yl} computed at $r^2$. With this, the proof of Theorem~\ref{theorem1} follows directly from Lemma~\ref{lemma2}. 

\vspace{-2mm}
\section{Numerical Computation Method for the CDF of a Generalized  Chi-Squared Random Variable}
\label{App:Imholf}

There is no closed-form expression for the \ac{PDF}, \ac{CDF}, and inverse \ac{CDF} of a generalized chi-squared variable $ Z = \sum_{i=1}^{N} \omega_i W_i$, for $W_i \sim \chi^2(k_i, \theta_i)$, where $\chi^2(k_i, \theta_i)$ denotes a noncentral chi-square distribution with $k_i$ degrees of freedom and noncentral parameter $\theta_i$. (Setting $k_i = 1$ and $\theta_i = \nu_i^2$ yields \eqref{eq:weightedY}.) Numerical methods can be used instead. The Imholf method computes $F_Z(z)$, the \ac{CDF} of $Z$, as a numerical integral $F_Z(z) \approx \frac{1}{2} - \frac{1}{\pi} \int_0^{U} \frac{\sin \beta(u,z)}{u \kappa(u)} \mathrm{d} u$, where 
\begin{subequations}
\begin{align}
   & \beta(u,z) =  \frac{1}{2} \sum_{i=1}^{N} \left( k_i \arctan (\omega_i u) + \frac{\theta_i \omega_i u}{1 + \omega_i^2 u^2}\right) - \frac{1}{2} z u ,\label{eq:CDF_imhof2}\\
   & \kappa(u) = \exp \left(\frac{1}{4} \sum_{i=1}^{N} k_i \ln(1 + \omega_i^2 u^2) + \frac{1}{2} \sum_{i=1}^{N} \frac{\theta_i \omega_i^2 u^2}{1 + \omega_i^2 u^2}   \right). \label{eq:CDF_imhof3}
\end{align}
\end{subequations}
The error for terminating at $U$ is given by $ \xi(U) = \frac{1}{\pi} \int_U^{\infty} \frac{\sin \beta(u,y)}{u \kappa(u)} \mathrm{d} u$. It is shown in \cite{imhof1961computing} that $ |\xi(U)| \leq \Xi(U)$, 
where
\begin{align}\label{eq:TU}
   \Xi(U) = \Big[ \pi K U^K \prod_{i=1}^{N} |\omega_i|^{\frac{k_i}{2}} \exp \big( \frac{1}{2} \sum_{i=1}^{n} \frac{\theta_i \omega_i^2 U^2}{1 + \omega_i^2 U^2}\big) \Big]^{-1}
\end{align}
with $K = \frac{1}{2} \sum_{i=1}^{N} k_i$. Thus, for any required accuracy $\epsilon$, a sufficiently large $U$ can be found to ensure $\Xi(U)\leq \epsilon$. 

\end{appendices}
\balance 

\bibliographystyle{IEEEtran}
\bibliography{ref}

\end{document}